\documentclass[revtex4]{emulateapj}

\usepackage{amsmath}
\usepackage{pdflscape}

\shortauthors{Sheehan et al.}
\shorttitle{Faint Compact Radio Sources in the ONC}

\begin{document}

\title{A VLA Survey For Faint Compact Radio Sources in the Orion Nebula Cluster}
\author{Patrick D. Sheehan$^1$, Josh A. Eisner$^1$, Rita K. Mann$^2$, and Jonathan P. Williams$^3$}
\affil{$^1$Steward Observatory, University of Arizona, 933 N. Cherry Avenue, Tucson, AZ, 85721}
\affil{$^2$National Research Council Canada, 5071 West Saanich Road, Victoria, BC, V9E 2E7, Canada}
\affil{$^3$Institute for Astronomy, University of Hawaii at Manoa, Honolulu, HI 96822}

\email{psheehan@email.arizona.edu}

\begin{abstract}
We present Karl G. Janksy Very Large Array (VLA) 1.3 cm, 3.6 cm, and 6 cm continuum maps of compact radio sources in the Orion Nebular Cluster. We mosaicked 34 square arcminutes at 1.3 cm, 70 square arcminutes at 3.6 cm and 109 square arcminutes at 6 cm, containing 778 near-infrared detected YSOs and $~190$ {\it HST}-identified proplyds (with significant overlap between those characterizations). We detected radio emission from 175 compact radio sources in the ONC, including 26 sources that were detected for the first time at these wavelengths. For each detected source we fit a simple free-free and dust emission model to characterize the radio emission. We extrapolate the free-free emission spectrum model for each source to ALMA bands to illustrate how these measurements could be used to correctly measure protoplanetary disk dust masses from sub-millimeter flux measurements. Finally, we compare the fluxes measured in this survey with previously measured fluxes for our targets, as well as four separate epochs of 1.3 cm data, to search for and quantify variability of our sources. 
\end{abstract}

\section{Introduction}

The Orion Nebular Cluster (ONC) presents an excellent example of star formation in a richly clustered environment, typical of star formation in our galaxy. Near-infrared surveys of the ONC find $>$700 YSOs, most of which are likely to harbor protoplanetary disks \citep{Hillenbrand2000}. {\it Hubble Space Telescope} ({\it HST}) images of the ONC also reveal ionized disks and dusty disks sillhoutetted against the backdrop of nebular emission \citep[e.g.,][]{ODell1994,Bally1998a,Smith2005,Ricci2008}. 

The O6 star $\theta^{1}$ Ori C, located in the central Trapezium Cluster, produces intense UV radiation that photoevaporates many of the nearby protoplanetary disks. The hot gas ionized by this intense radiation expands freely and flows away at the local sound speed into lower pressure regions \citep[e.g.,][]{Henney1998}. The ionized winds from the protoplanetary disks emit strong free-free emission at radio wavelengths \citep[e.g.,][]{Garay1987,Churchwell1987}.

Compact radio sources have long been known in the ONC \citep[e.g.,][]{Moran1982,Garay1987,Churchwell1987,Felli1993a,Zapata2004a,Zapata2004b}. They were first identified as free-free emission by \citet{Garay1987}, and suggested to be the ionized material evaporated from protostellar disks by \citet{Churchwell1987}. Observations of the ONC with the {\it Hubble Space Telescope} firmly established these compact structures as externally ionized protoplanetary disks \citep[e.g.,][]{ODell1993}.

Measurements of the masses of protoplanetary disks are crucial for understanding evolution, as well as potential for planet formation. Disk mass measurements are typically made by observing dust continuum emission at long wavelengths, where the emission is optically thin and probes the entirety of the disk \citep[e.g.,][]{Beckwith1990}. Towards this end, a host of millimeter interferometric surveys of the ONC have previously been carried out \citep[e.g][]{Mundy1995,Bally1998b,Williams2005,Eisner2006,Eisner2008,Mann2009,Mann2010,Mann2014}. 

\tabletypesize{\scriptsize}
\begin{deluxetable*}{ccccccccccc}
\tablehead{\colhead{Band} & \colhead{Configuration} & \colhead{Date} & \colhead{Int. Time} & \colhead{RMS} & \colhead{Peak RMS} & \colhead{Beam} & \colhead{No. Beams} & \colhead{Total} & \colhead{$>6\sigma$} & \colhead{$>4.5\sigma$}\\ \colhead{ } & \colhead{ } & \colhead{ } & \colhead{[min]} & \colhead{[$\mu$Jy]} & \colhead{[$\mu$Jy]} & \colhead{ } & \colhead{ } & \colhead{Detections} & \colhead{Detections} & \colhead{Detections}}
\startdata
1.3 cm & B & Nov. 10, 2013 & 62 & 33 & $\sim93$ & 0.33"$\times$0.21" & 1.5$\times10^6$ & 79 & 57 & 22 \\
6 cm & A & Mar. 3, 2014 & 7 & 37 & $\sim150$ & 0.40"$\times$0.28" & 3.1$\times10^6$ & 108 & 87 & 21 \\
3.6 cm & A & Mar. 3, 2014 & 9.5 & 30 & $\sim70$ & 0.24"$\times$0.18" & 5.1$\times10^6$ & 98 & 80 & 18 \\
1.3 cm & A & Mar. 3, 2014 & 49 & 25 & $\sim50$ & 0.09"$\times$0.08" & 14.9$\times10^6$ & 70 & 54 & 16 \\
1.3 cm & A & Mar. 7, 2014 & 49 & 26 & $\sim100$ & 0.08"$\times$0.08" & 15.7$\times10^6$ & 73 & 56 & 17 \\
1.3 cm & A & May 3, 2014 & 36.5 & 22 & $\sim85$ & 0.10"$\times$0.07" & 14.6$\times10^6$ & 89 & 67 & 22 \\
1.3 cm & A \& B combined & \nodata & \nodata & 12 & $\sim50$ & 0.09"$\times$0.09" & 12.4$\times10^6$ & 126 & 98 & 28
\enddata
\label{table:obs}
\tablenotetext{a}{$>6\sigma$ Detections refers to the number of sources detected in each map because they pass the blind detection threshold. $>4.5\sigma$ Detections refers to the additional sources detected in a catalog driven search, and Total Detections is the sum of those numbers.}
\end{deluxetable*}

These surveys are complicated by potential contamination of the millimeter dust continuum emission by free-free emission from ionized disk winds. Disk mass measurements are facilitated at shorter wavelengths, of 1.3 mm or 870 $\mu$m, where the ratio of dust emission to free-free emission is expected to be more favorable. Even here, however, free-free emission can contribute significantly to the observed brightnesses of the sources \citep[e.g.,][]{Eisner2008,Mann2009,Mann2010,Mann2014}. 

Observations at longer radio wavelengths can help to constrain the free-free contribution at shorter wavelengths. Free-free emission has a flat spectrum ($F_{\nu} \propto \nu^{-0.1}$) when optically thin, as is expected to be true at millimeter and centimeter wavelengths \citep[e.g.][]{Eisner2008,Mann2009,Mann2010,Mann2014}. Optically thick free-free emission can span a range of spectral indices, but the emission usually only becomes optically thick at wavelengths longer than $\sim10$ cm \citep[e.g.][]{Eisner2008}. Dust emission, however, has a steep spectral index ($F_{\nu} \propto \nu^{2 + \beta}$, $\beta = 0-2$) which falls off rapidly at longer wavelengths. Free-free emission can therefore be constrained at longer radio wavelengths where the contribution from dust emission to the flux is small. Radio fluxes may also in some cases be affected by magnetospheric flaring from young stars, exhibiting gyrosynchrotron emission with a steep negative spectral index when optically thin \citep[$F_{\nu} \propto \nu^{-0.7}$; e.g.][]{Feigelson1999,Rivilla2015}, or a steep positive spectral index when optically thick at lower frequencies ($F_{\nu} \propto \nu^{2.5}$).

Previous studies have used the VLA to search for compact radio sources in the ONC \citep[e.g.,][]{Felli1993a,Zapata2004a}, and fluxes produced by those studies have been used to correct for free-free contamination in disk mass studies \citep[e.g.,][]{Eisner2008,Mann2010,Mann2014}. The expanded capabilities of the VLA correlator \citep{Perley2009}, now enable surveys of much higher sensitivity than were previously possible. More recent surveys have taken advantage of this increase in sensitivity to map star forming regions, including the ONC at 4.5 GHz and 7.5 GHz \citep{Dzib2013,Kounkel2014,Forbrich2016}. This enhanced sensitivity is well-matched to the deeper observations now enabled with ALMA.

Here we present new high resolution Karl G. Jansky Very Large Array (henceforth JVLA to avoid confusion with previous surveys using the original VLA) maps of the ONC at 1.3 cm, 3.6 cm, and 6 cm to study the free-free emission from ONC cluster members. In Section 2 we describe our observations and maps of the ONC. In Section 3 we detail our methodology for searching for compact radio sources, as well as our model for characterizing the free-free emission. In Section 4 we compare our results to previous catalogs of compact radio sources in the ONC, discuss the nature of the sources we detect, and show that our measurements are crucial for accurately measuring disk masses of protoplanetary disks from both current and future submillimeter surveys.

\section{Observations \& Data Reduction}

We imaged the Orion Nebula Cluster in 1.3 cm, 3.6 cm, and 6 cm wavelength continuum emission with the JVLA between November 2013 and May 2014. The 3.6 cm and 6 cm maps were observed using the `A' configuration (baselines ranging from 680 m to 36 km), and the 1.3 cm data were taken in three epochs with the `A' configuration and one epoch with the `B' configuration (baselines ranging from 210 m to 11 km). Details of the observations and maps are provided in Table \ref{table:obs}.

The 3.6 cm and 6 cm data were taken simultaneously in 32 128 MHz bands, split evenly between 3.6 cm and 6 cm. Each band contained 64 2 MHz channels, and the bands were arranged continuously from 4.488 - 6.512 GHz at 6 cm and from 8.116 - 10.012 GHz at 3.6 cm, for a total of 2 GHz of continuum bandwidth each. 

The field of view of the JVLA antenna primary beam at 6 cm, FWHM of 9', encompasses all 778 YSOs from \citet{Hillenbrand2000}, and 196 of the 196 HST detected proplyds \citep{Ricci2008}. 141 of the 196 HST detected proplyds are also detected as sources in \citet{Hillenbrand2000}. We therefore use a single pointing to image the field at 6 cm. At 3.6 cm the field of view is 5$'$, so we imaged the field with two pointings that encompassed 778 YSOs and 187 {\it HST}-detected proplyds.

For a rectangular mosaic the Nyquist sampling theorem suggests that a pointing spacing of FWHM/2 or better is needed \citep[e.g.][]{Cornwell1988}, but since we are interested in compact sources, Nyquist sampling is not crucial \citep[e.g.][]{Eisner2008}. At 3.6 cm the FWHM/2 is between 2.1' and 2.6' across the band. The two 3.6 cm pointings are separated by 2.4', so the map is sub-Nyquist sampled at the low frequency end of the band, but not at the high frequency end of the band.

The 1.3 cm data were taken in 64 128 MHz bands arranged from 17.976 - 26.024 GHz. Each band was composed of 64 2 MHz channels, for a total of 8 GHz of bandwidth. Most of the data, however, from 17.976 - 22.024 GHz is affected by significant RFI, so we exclude that data from our analysis. The 1.3 cm data therefore has an effective bandwidth of 4 GHz.

\begin{figure*}
\centering
\includegraphics[width=7in]{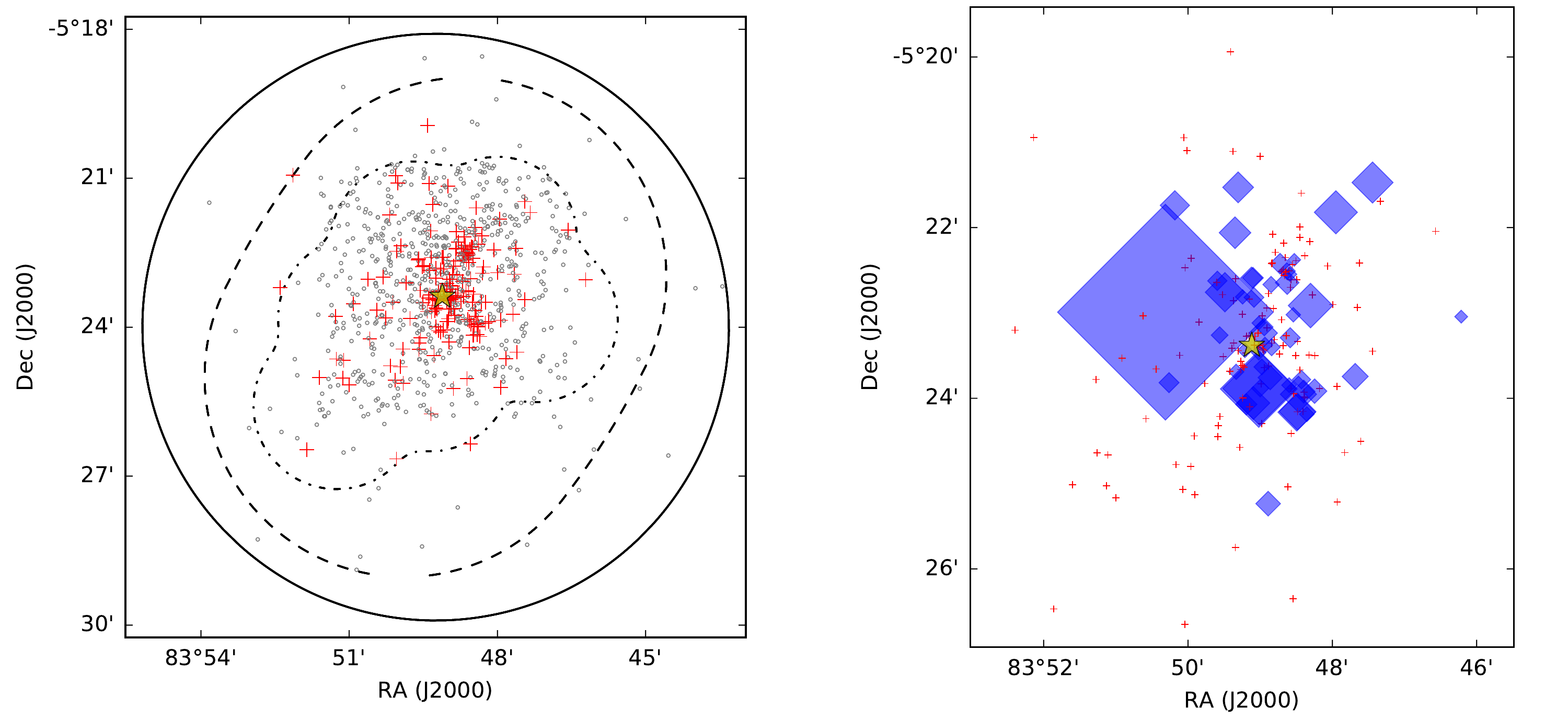}
\caption{The fields we image, out to the 20\% gain contour at 6 cm (solid) and the 10\% gain contour at 3.6 cm (dashed) and 1.3 cm (dash-dotted) observations, with a yellow star representing the location of $\theta^1$ Ori C. On the left we show all of the sources we detected in at least one of our bands with red plusses and the known sources surveyed but not detected with grey circles. On the right we show sources found to be variable with blue rectangles whose size is proportional to how variable the source is. The largest symbols represent a variability amplitude of 900\% while the smallest represent an amplitude of 20\%. The detected sources which are not variable are shown again with red plusses.}
\label{fig:source_map}
\end{figure*}

A field of view containing 778 YSOs and 193 {\it HST} detected proplyds was mosaicked using 7 pointings. A two dimensional map is Nyquist sampled if the pointing spacing is FWHM/$\sqrt3$ or better, but since we are interested here in compact sources, Nyquist sampling is, again, not crucial. At 1.3 cm FWHM/$\sqrt3$ is between 1.2' and 1.4' across the band. The mosaic spacings range between 1-2', so the map is largely not Nyquist sampled. We show the field of view of our observations for each band in Figure \ref{fig:source_map}.

The data were calibrated and imaged using the CASA software package. Antenna-based complex gains were calculated using periodic observations of the quasar J0541-0541. Bandpass solutions for each antenna were calculated from observations of the quasar J0319+4130, and the overall flux density scale was calculated using models included in CASA for 3C48.

We produced maps of the ONC at each frequency by Fourier transforming the complex visibilities, using the mosaicking modes for the 1.3 cm and 3.6 cm maps. We weighted the data with a robust parameter of 0, which provided a good balance between the high sensitivity of normal weighting and the high spatial resolution of uniform weighting. We also used the multi-frequency synthesis option with nterms=1 for no source frequency dependence \citep{Rau2011}. Our goal is to search for compact structures in the Orion Nebula, so we removed baselines shorter than 100 k$\lambda$ from our data before inverting the visibilities. The spatial scales eliminated by this cut correspond to structures greater than 2$''$, meaning that large scale structure from the Orion Nebula has been resolved out of our maps. For these observations our reference frequencies are 5.5 GHz for the 6 cm map, 9 GHz for the 3.6 cm map, and 22.5 GHz for the 1.3 cm map. We image the 6 cm data out to the 20\% gain contour at 5.5 GHz, and the smaller 3.6 cm and 1.3 cm maps out to the 10\% gain contour at 9 GHz and 22 GHz respectively\footnotemark. We imaged each 1.3 cm epoch separately to study the variability of the bright sources, and together to increase our sensitivity to look for faint sources in the map.

\footnotetext{These correspond to the 33 and 10\% gain contours for the low and high frequency 6 cm band edges respectively, the 16 and 6\% gain contours for the 3.6 cm band edges, and the 14 and 6\% gain contours for the 1.3 cm band edges.}

We CLEANed the images using the Clark algorithm \citep{Clark1980}. Sources above $10\sigma$ were initially identified for CLEANing by visual inspection. The maps were CLEANed down to the rms, as measured in source-free regions of the maps, listed in Table \ref{table:obs}. Post-source detection, we could re-CLEAN the image using the new detections, however the sidelobes of these sources are low enough to be below the noise level, and the improvement by CLEANing them is minimal and the computational requirements are significant. 

We used a single iteration of self-calibration on the 6 cm data, correcting for just the phases of our data from a model produced by an initial CLEANing of the data. This improved the rms ($\sim50$ $\mu$Jy to $\sim40$ $\mu$Jy) in crowded regions of the map or near bright sources with significant beam artifacts.

We self-calibrated the data using a model produced from both fields simultaneously. We find that self-calibrating the fields separately and then imaging them jointly produced ringing in the image that was removed by self-calibrating the data together. We used two iterations of self-calibration, first solving for the phases from our initial model, and then solving for the amplitudes and any residual phase errors in a second iteration. We apply amplitude self-calibration because it helps to remove residual artifacts around bright sources in our map. It does not change the flux in our maps markedly. This improved the rms from $\sim80$ $\mu$Jy near bright sources with significant beam artifacts to $\sim45$ $\mu$Jy.

We self-calibrated fields including the brightest sources together, which is necessary to remove ringing like in the 3.6 cm maps, using a single iteration of self-calibration to correct phase errors in the data. The self-calibration improved the rms by as much as a factor of 3 near bright sources with significant beam artifacts (e.g. $\sim50$ $\mu$Jy to $\sim20$ $\mu$Jy for the combined 1.3 cm map, $\sim100$ $\mu$Jy to $\sim35$ $\mu$Jy for the 1.3 cm data taken on March 3, 2014).

After CLEANing, each map was corrected for attenuation by the primary beam, using the primary beam at the central frequency of each band. The bandwidth of our observations is a significant fraction of the central frequency, however, so the primary beam correction may vary significantly over the band. We have computed the error induced in wideband fluxes measured when correcting by the primary beam of the central frequency, rather than the appropriate primary beam for each channel, and find that this error is $<$5\% for realistic spectral indices (-0.1-2).

Finally, we restored each map with a CLEAN beam whose size is determined by a Gaussian fit to the central peak of the dirty beam for that map. The size of this beam is given approximately by $\lambda/B_{max}$ for the map, but the exact size and shape depend on the distribution of baselines in the $uv$-plane and the choice of weighting function. We list the beam sizes for each map in Table \ref{table:obs}. After our initial CLEANing of the data we self-calibrated on the brightest sources in our maps to remove residual beam structure and improve the sensitivity, particularly in crowded regions. 

\begin{figure}[b]
\centering
\includegraphics[width=3in]{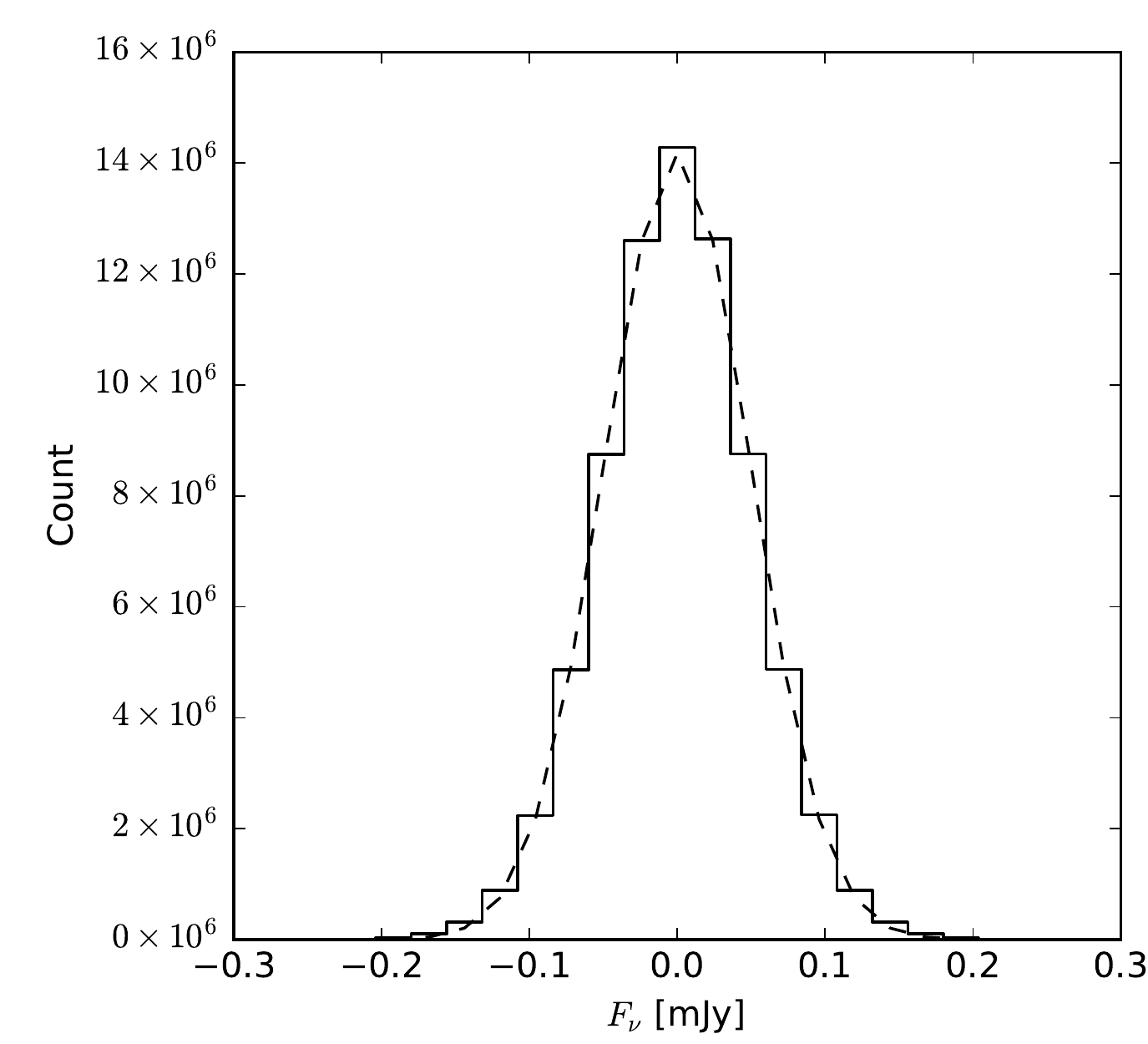}
\caption{We show a histogram of all the pixel values within the 50\% gain contour of our 6 cm residual map. We also show the best fit Gaussian to the distribution with the dashed line. Here we show only the 6 cm map, but we have produced similar figures for the 3.6 cm and 1.3 cm maps and find that both of those distributions are also Gaussian, so we can use a $\sigma$-cut to confidently distinguish between real sources and noise spikes in our images.}
\label{fig:noise}
\end{figure}

\section{Analysis}

\subsection{Source Detection}

\begin{figure*}
\centering
\includegraphics[width=4in]{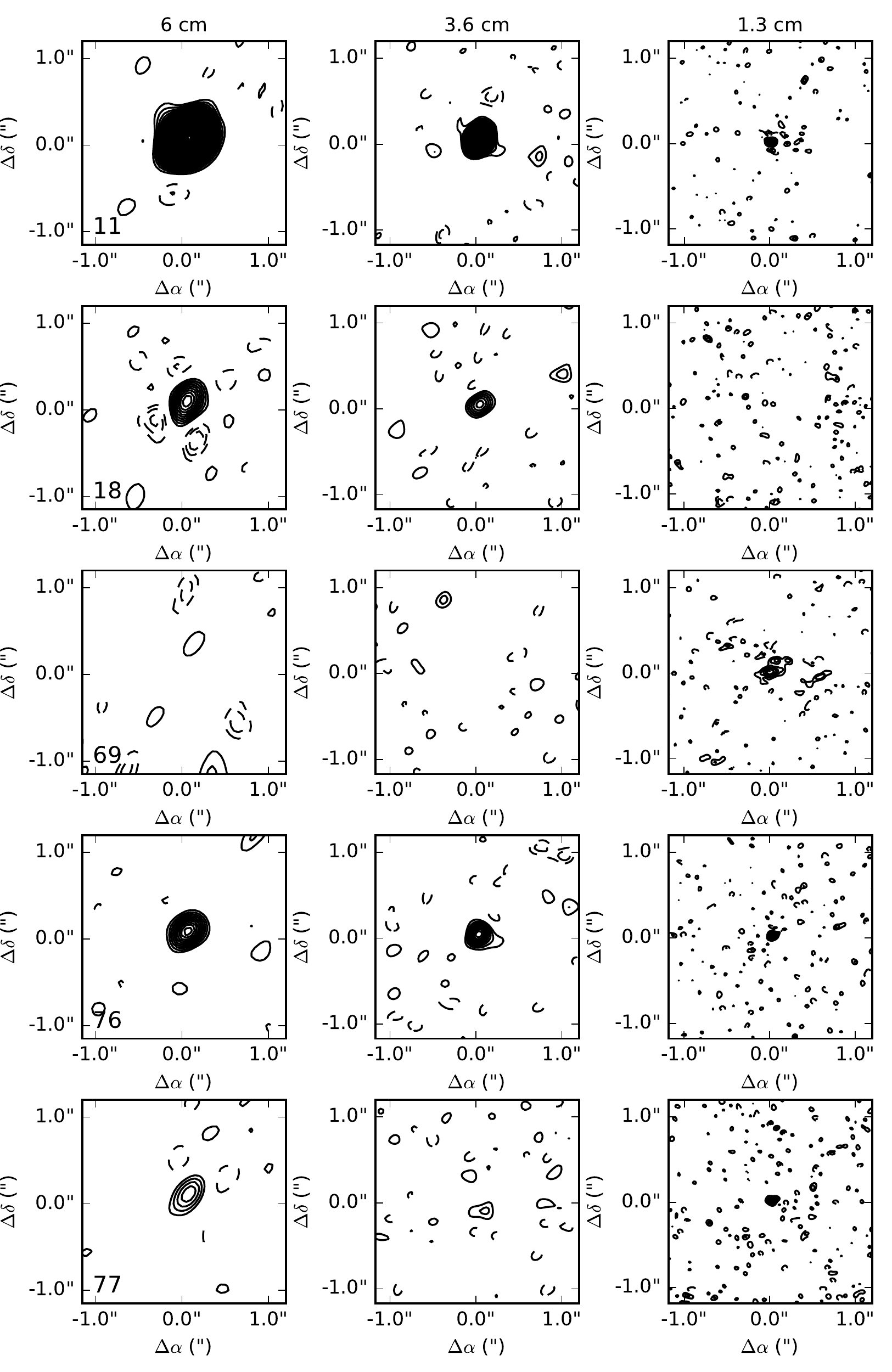}
\caption{Contour images of sources detected in our 6 cm, 3.6 cm or 1.3 cm continuum maps. Each row shows a single source in each band. At 1.3 cm we show only one epoch as a representative image of the source. Contour increments are $1\sigma$, beginning at $\pm2\sigma$, where $\sigma$ is determined locally for each object. This figure is continued at the end of the text.}
\label{fig:source_plots}
\end{figure*}

In each of our VLA maps we search for sources detected above a certain signal-to-noise threshold. Our maps contain $>10^6$ synthesized beams (see Table \ref{table:obs}), so we must employ a relatively conservative threshold to ensure that we do not select noise spikes in the images as real detections. The noise in each map follows a Gaussian distribution (see Figure \ref{fig:noise}), so we expect $\ll$1 noise spike to fall above a $6\sigma$ detection threshold. We therefore use $6\sigma$ as our detection limit.

We can also use catalogs of previously known source positions to target our search. We search our maps at the positions of $>$700 near-infrared detected sources \citep{Hillenbrand2000} and $\sim$200 $HST$ detected proplyds \citep[][with an overlap of about $\sim$150 of the near-infrared detected sources]{Ricci2008}. We also search the coordinates of known submillimeter sources detected with the SMA, CARMA, and ALMA that lack counterparts at infrared wavelengths \citep{Eisner2008,Mann2010,Mann2014}. Finally, we search for compact radio sources which were detected with the VLA by previous surveys \citep{Felli1993a,Kounkel2014}. Due to the smaller number of synthesized beams being probed ($\sim800$), we expect $\ll$1 noise spike to fall above a $4.5\sigma$ level. For each previously identified source we search for a detection above $4.5\sigma$ within a radius of 0.5", typical of the sizes of beams from these previous studies.

The rms at each pixel is calculated from a 128 by 128 pixel box surrounding that pixel in the residual map. The rms in the map is generally low ($\sim$25 $\mu$Jy in the 1.3 cm maps, $\sim30$ $\mu$Jy at 3.6 cm, and $\sim37$ $\mu$Jy at 6 cm). However, the central region of each map exhibits beam artifacts from a cluster of bright sources and poor sampling of large scale emission. The rms in these regions can be much higher than the rest of the map ($\sim100$ $\mu$Jy in the 1.3 cm maps, $\sim70$ $\mu$Jy at 3.6 cm, and $\sim150$ $\mu$Jy at 6 cm; see Table \ref{table:obs}).

We list the total number of sources detected in each map in Table \ref{table:obs}. For each map we also provide a breakdown of the number of sources detected in our blind search as well as the additional number of sources detected from the catalog driven search. We detect 108 objects in our 6 cm map, 98 objects in our 3.6 cm map, and a total of 144 objects across all of our 1.3 cm maps. In all we detect 175 distinct sources across all of our maps. We show the position of every detected source in our maps in the left panel of Figure \ref{fig:source_map}.

Of the 175 unique compact radio sources, 120 sources are associated with YSOs detected in near-infrared surveys \citep[e.g.][]{Hillenbrand2000}, and 67 sources are associated with {\it HST} detected proplyds. 149 have previous radio detections, and 40 have been previously detected at millimeter wavelengths. We also report the detection of 11 sources here for the first time at any wavelength.

We fit every detected source with a two dimensional Gaussian to determine position, extent, and total source flux. For sources identified in the previously mentioned catalogs that are not detected in our maps, we also integrate over a 1" aperture centered on the known source position to produce an unbiased estimate of the signal (or noise) towards that position. We include a 10\% error on the measurement to account for systematic errors in the band-to-band flux calibration. These intensities, measured towards all cataloged objects in our field of view, are presented in Table \ref{table:fluxes}. The print version of this paper presents only the first page of that table. We also plot images of those sources in Figure \ref{fig:source_plots}. 

Our catalog of sources, as presented in Table \ref{table:fluxes} is sorted by right ascension and then given a catalog `ID', which we list in Table \ref{table:fluxes}. We refer to each source by this ID throughout the remainder of the text and figures. In Table \ref{table:fluxes} we also list the proplyd name, identification from early ONC radio surveys \citep[e.g.,][]{Garay1987,Felli1993b}, identification from \citet{Zapata2004a}, or identification from \citet{Hillenbrand2000} when applicable. 

\begin{figure*}
\centering
\includegraphics[width=7in]{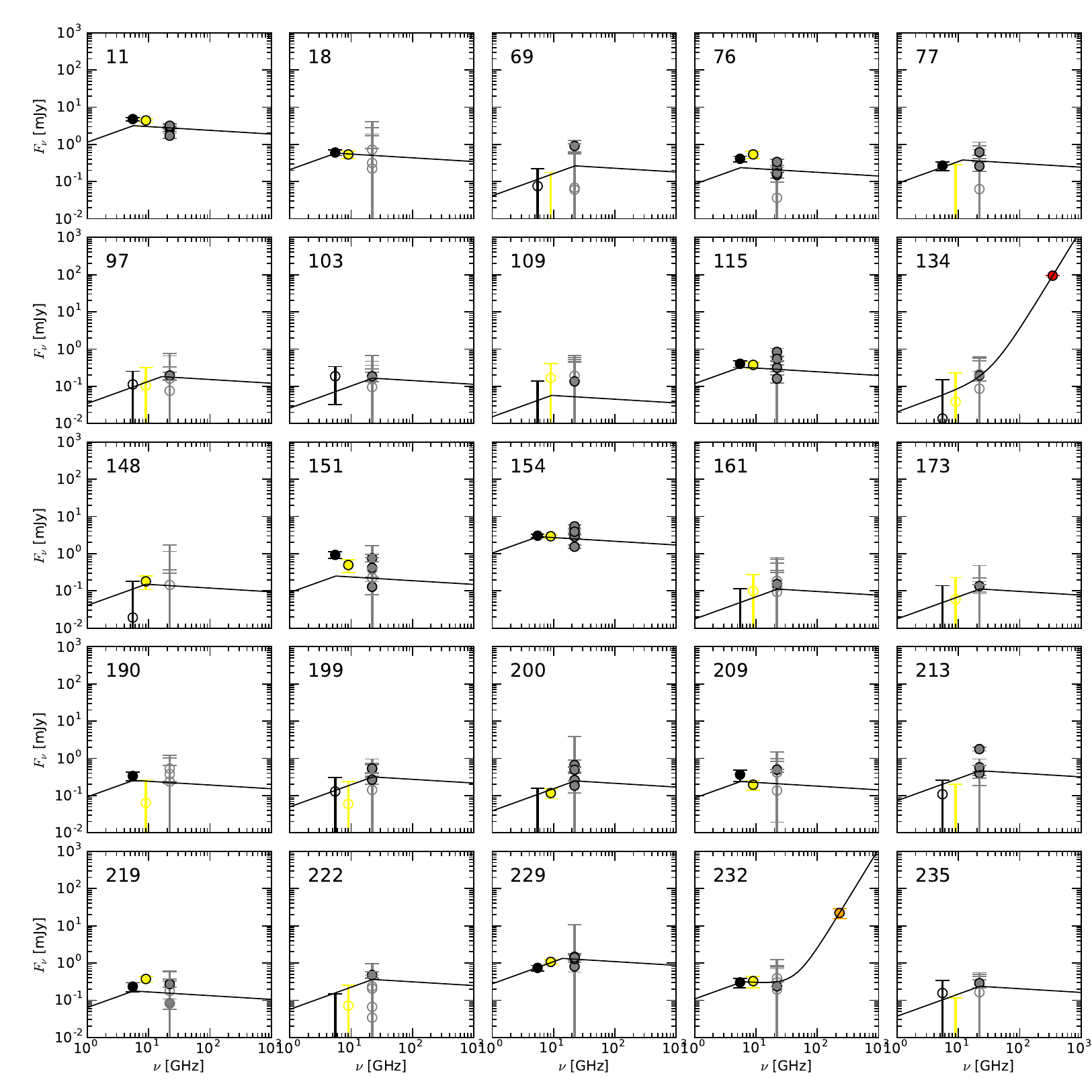}
\caption{The millimeter and radio SEDs for all of the sources detected in our maps. We also show the best fit dust + free-free emission model for each source, as described in Section 3.2. Black, yellow and grey points are the 6 cm, 3.6 cm, and 1.3 cm flux measurements for objects detected in our maps. Circles with colored faces indicate that the source was detected by our search routines, while open face circles are fluxes measured in an aperture around a known source position. Orange data points are 3mm, 1.3 mm, and 870 $\mu$m fluxes from \citet[][and references therein]{Eisner2008}. Green data points are 870 $\mu$m fluxes from \citet{Mann2010}, and red data points are 870 $\mu$m fluxes from \citet{Mann2014}. The fluxes shown here are all measured with one of the SMA, CARMA, ALMA, OVRO, or the VLA. This figure is continued at the end of the text.}
\label{fig:sed_fits}
\end{figure*}

\subsection{Estimating the Free-Free Emission Spectrum}

\begin{figure*}
\centering
\includegraphics[width=7in]{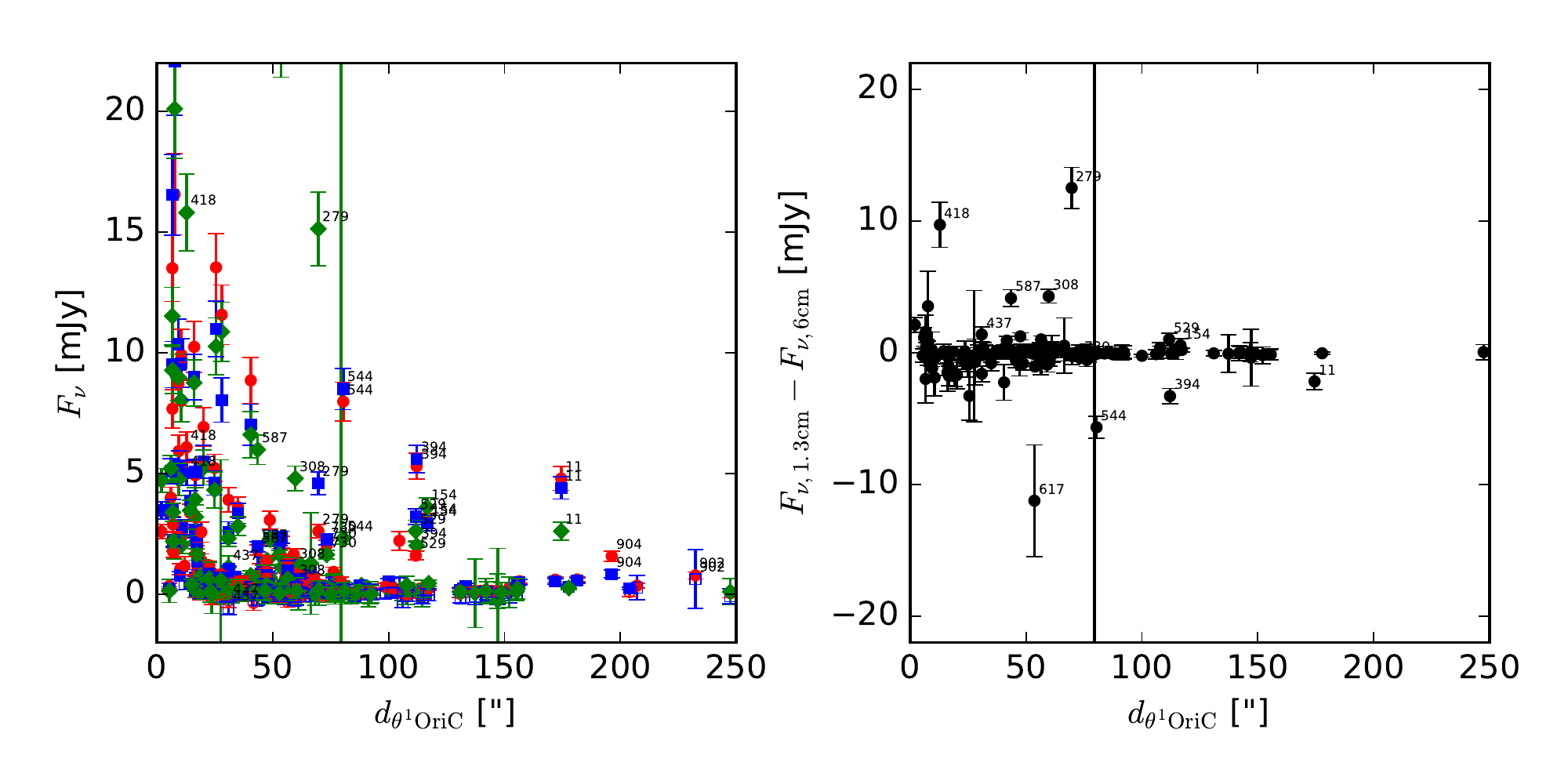}
\caption{(left) The measured radio flux of each of our detected objects at 1.3 cm (green diamonds), 3.6 cm (blue squares), and 6 cm (red circles) as a function of projected distance from $\theta^1$ Ori C. (right) Difference in measured 1.3 cm and 6 cm fluxes as a function of projected distance from $\theta^1$ Ori C. With the exception of a few outliers, we find that radio fluxes for our targets decrease with increasing projected separation, as we would expect for free-free emission driven by the powerful ionizing radiation of $\theta^1$ Ori C. This is also consistent with the difference in 1.3 cm and 6 cm fluxes, which falls near zero for most sources. Optically thin free-free emission is expected to have a roughly flat spectrum at these wavelengths, so we would expect the differences in those measurements to fall near zero.We label the significant outliers with the source ID for reference in future sections.}
\label{fig:radial_fluxes}
\end{figure*}

Evidence suggests that the proplyds are undergoing mass loss from photoevaporation by the nearby O star $\theta^1$ Ori C, so the free-free emission we detect here is likely due to a wind \citep[e.g.,][]{Churchwell1987,Henney1998}. For emission from a spherically symmetric wind with an arbitrary $n \propto r^{-\alpha}$ density profile the expected spectral dependence of free-free emission is
\begin{equation}
F_{\nu,ff} = 
\begin{cases}
F_{\nu,turn} \left(\frac{\nu}{\nu_{turn}}\right)^{-0.1} & \nu \geq \nu_{turn} \\
F_{\nu,turn} \left(\frac{\nu}{\nu_{turn}}\right)^{(4\alpha-6.2)/(2\alpha-1)} & \nu < \nu_{turn}
\end{cases}
\end{equation}
\citep{Wright1975}. $\nu_{turn}$ is the frequency where the wind becomes partially optically thick, and is determined by the radius of the inner boundary of the ionized envelope. High turnover frequencies indicate more compact inner boundaries. When the wind becomes fully optically thick at very low frequencies the spectrum follows the typical $F_{\nu} \propto \nu^2$ spectrum expected for optically thick thermal emission. 

For a fully ionized wind with a constant mass-loss rate we expect $\alpha=2$ and the spectral dependence of free-free emission is
\begin{equation}
F_{\nu,ff} = 
\begin{cases}
F_{\nu,turn} \left(\frac{\nu}{\nu_{turn}}\right)^{-0.1} & \nu \geq \nu_{turn} \\
F_{\nu,turn} \left(\frac{\nu}{\nu_{turn}}\right)^{0.6} & \nu < \nu_{turn}
\end{cases}.
\end{equation}
Steeper density profiles may lead to steeper spectral dependences below the turnover frequency \citep[e.g.,][]{Plambeck1995}. Here, for simplicity, we adopt the solution for a fully ionized wind with a constant mass-loss rate. Many of our sources show evidence for a free-free turnover (see Figure \ref{fig:sed_fits}), so we adopt a model including a turnover in the spectrum.

At higher frequencies, dust emission is expected to dominate. The differences in the expected spectral slopes between dust and free-free emission allows us to characterize each separately by observing our targets at a range of wavelengths. For each of our detected sources we fit a simple model to the known radio, millimeter, and sub-millimeter photometry:
\begin{equation}
F_{\nu} = F_{\nu,ff} + F_{\nu,dust,230 GHz} \left(\frac{\nu}{230 \text{GHz}}\right)^{2+\beta}.
\end{equation}
Here we assume $\beta = 0.7$, consistent with previous studies of protoplanetary disks in other star forming regions \citep[e.g.,][]{Rodmann2006,Ricci2010b,Ricci2010a}.

We fit the SED of each source by searching a grid over a large range of parameter space of $\nu_{turn}$, $F_{\nu,turn}$, and $F_{\nu,dust,230GHz}$ for a minimum in $\chi^2$. We then use a second, finely spaced, grid search based on the initial search to find the best $\chi^2$ fit.

$\nu_{turn}$, $F_{\nu,turn}$ and $F_{\nu,dust,230 GHz}$ are left as free parameters in the grid search. If a source has no submillimeter detections \citep[$\geq90$ GHz;][]{Eisner2008,Mann2010,Mann2014}, we assume that $F_{\nu,dust,230 GHz} = 0$. In that case we also require $5.5 \text{GHz} \leq \nu_{turn} \leq 22 \text{GHz}$, because outside of this range we cannot constrain $\nu_{turn}$. If a source does have sbmillimeter detections we only require $5.5 \text{GHz} \leq \nu_{turn}$.

Sources 281, 391, 416, 423, 430, 433, 442, 512, 516, 537, 564, and 595 are all extended sources that are marginally resolved by our 3.6 cm and 6 cm maps, as well as in our B-configuration 1.3 cm observations. In our A-configuration 1.3 cm observations, however, these sources are very well resolved. In fact, they are so well resolved that much or all of the emission from the source is resolved out. As such we exclude the A-configuration flux measurements from our SED fitting, as flux variations are likely due to structure being resolved out rather than actual variability.

Some of our sources are variable across our multiple epochs of 1.3 cm data (see Section 4.2). We account for this variability in our modeling by including the measured flux at each epoch and allowing the variability to influence the uncertainty of our parameter estimation. Sources that are more variable will also have more uncertainty in model fits.

\begin{figure*}
\centering
\includegraphics[width=7in]{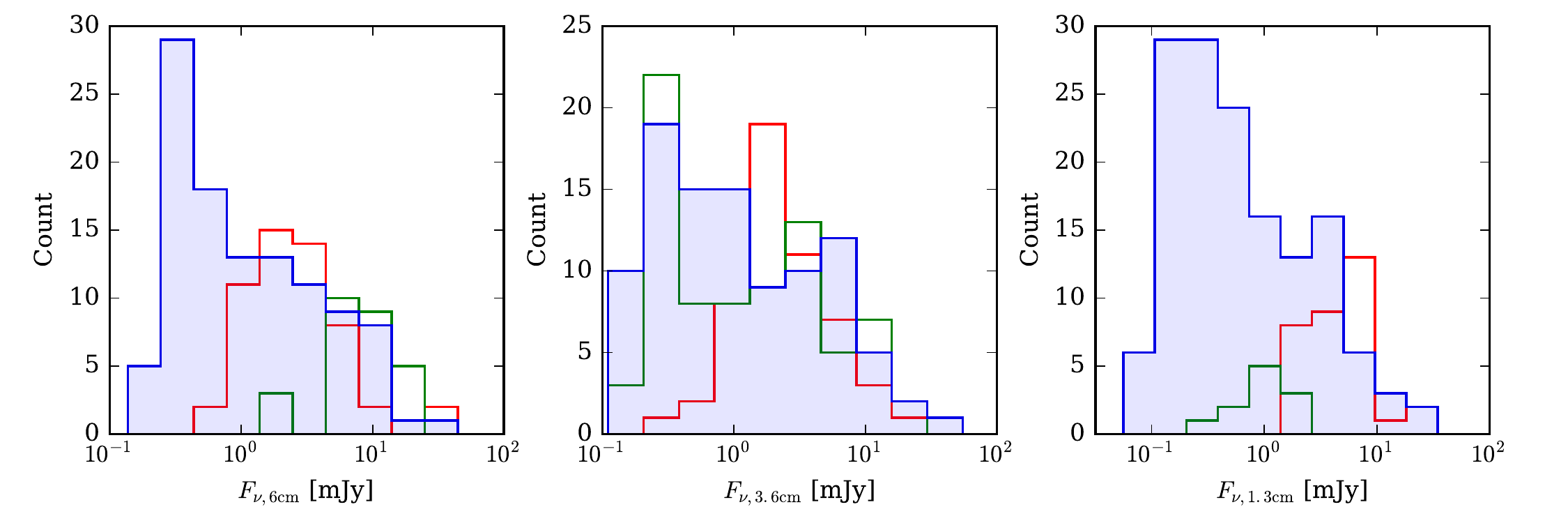}
\caption{Histograms of the fluxes of sources detected in each of our maps. We also show a histogram of the compact radio sources detected in previous studies. Blue shows the histogram of detected sources from this work. Green shows the histograms of detected sources from \citet{Felli1993b} (6 cm), \citet{Zapata2004a} (3.6 cm), and \citet{Zapata2004b} (1.3 cm). Red shows the 4.5 GHz (6 cm) and 7.5 GHz (3.6 cm) detections from \citet{Kounkel2014}, and the 2 cm detections from \citet{Felli1993a}. We do not show the \citet{Forbrich2016} 6 cm sample, which includes 477 sources fainter than 0.3 mJy.}
\label{fig:flux_hist}
\end{figure*}

We list the parameters of our best fit models to each source detected in our maps in Table \ref{table:modelparams}. The photometry, along with the best fit model, for each source is plotted in Figure \ref{fig:sed_fits}.

The origin of the radiation ionizing these sources has been the subject of much debate. Early radio studies of the region disagreed as to whether these sources were externally ionized by radiation from $\theta^1$ Ori C or ionized internally by a young massive star, and as to whether these objects are dense neutral condensations or protoplanetary disks \citep[e.g.,][]{Garay1987,Churchwell1987}, although these studies are complicated by the fact that only projected, and not actual, distances from $\theta^1$ Ori C are known. Since these early studies, {\it HST} imaging \citep[e.g.,][]{ODell1993} and detailed modeling of those images \citep[e.g.][]{Henney1998} has favored protoplanetary disks ionized by $\theta^1$ Ori C.

Free-free emission powered by ionizing radiation from $\theta^1$ Ori C should decrease with increasing separation from $\theta^1$ Ori C. We show the measured flux versus distance in the left panel of Figure \ref{fig:radial_fluxes}. For most sources there is a trend of decreasing radio flux with increasing separation, suggesting that they are exhibiting free-free emission from gas ionized by $\theta^1$ Ori C. We also find that the difference in their 1.3 cm and 6 cm flux is close to zero, as expected for optically thin free-free emission, which has a roughly flat spectrum (see the right panel of Figure \ref{fig:radial_fluxes}). We note that here we report projected distances. Actual separations are greater than or equal to this number. We discuss the outliers of these trends below, in Section 4.3.

For this study, we are only concerned with whether these sources are emitting thermal free-free emission or not so that we can characterize the emission and remove it from dust emission for disk mass studies, but on the surface Figure 5 would seem to suggest that these sources are externally ionized by $\theta^1$ Ori C. The detailed structure of these compact objects is beyond the scope of this paper, as the radio emission can be well characterized without that knowledge, but we will revisit the subject in a more detailed study in the future.

\section{Discussion}

\subsection{Comparison with Previous Radio Surveys}

Many compact radio sources have previously been identified at a range of wavelengths in the ONC through VLA surveys of the region. The earliest searches for compact radio sources in the ONC were conducted primarily at 20 cm, 6 cm, 2 cm, and 1.3 cm \citep[e.g.,][]{Garay1987,Churchwell1987,Felli1993a,Felli1993b}. These surveys were state of the art at the time, with rms as low as 0.18 mJy beam$^{-1}$ at 2 cm \citep{Churchwell1987,Felli1993a}, 0.23 mJy beam$^{-1}$ at 6 cm \citep{Felli1993b}, or 1.0 mJy beam$^{-1}$ at 1.3 cm \citep{Garay1987}. These searches identified 49 compact radio sources in the ONC.

A more recent survey mapped a 4' $\times$ 4' region of the the ONC at 3.6 cm using the VLA. This survey achieved a sensitivity of 0.03 mJy beam$^{-1}$ and uncovered 77 compact radio sources \citep{Zapata2004a}. Of these 77 sources, 38 were previously known from the earlier studies mentioned above, while 39 were new centimeter detections. \citet{Zapata2004b} also mapped a 30" $\times$ 30" region in OMC-1 South at 1.3 cm with the VLA. They achieved an rms of 0.07 mJy beam$^{-1}$, but due to the limited area of their maps only detected 11 sources.

A recent survey mapped out a large region encompassing $\lambda$ Ori, Lynds 1622, NGC 2068, NGC 2071, NGC 2023, NGC 2024, $\sigma$ Ori, the ONC, and Lynds 1641 with the VLA at 4.5 GHz and 7.5 GHz with a 60 $\mu$Jy sensitivity \citep{Kounkel2014}. They found $>350$ sources over the area of their map, 54 of which overlap with the area we survey. The majority of their detected sources also have spectral indices consistent with flat spectra.

Here we compare these previous surveys with our own JVLA maps. In Figure \ref{fig:flux_hist} we plot the distribution of fluxes for compact sources detected in our maps as well as the distribution of fluxes for previously identified compact radio sources at the same wavelength. The most extensive existing studies at 1.3 cm are limited by either survey area or sensitivity so we also compare our 1.3 cm detections with previous 2 cm detections.

Of the 49 compact radio sources detected by initial surveys \citep[e.g.][]{Garay1987,Churchwell1987,Felli1993a,Felli1993b}, we have detected 37 in our maps. We have also recovered 64 of the 77 sources detected by \citet{Zapata2004a}, 9 of the 11 sources found by \citet{Zapata2004b}, 42 of the 54 sources found by \citet{Kounkel2014}, and 144 of the 556 sources found by \citet{Forbrich2016}. We detect 29 of the 35 sources that were previously detected at 2 cm. We also report the detection of 135 sources that have not previously been detected at 1.3 cm, 34 at 3.6 cm, 4 at 6 cm, and 26 sources that have not previously been detected at any radio wavelengths. The sources that were previously detected, but that we do not detect in our maps, are likely variable given the deeper sensitivity in our JVLA data.

\subsection{Variability}

Previous radio studies of the ONC explored multiple epochs of data to search for evidence of source variability. \citet{Felli1993b} monitored the ONC at 2 cm and 6 cm for a period of 7 months and found 13 sources to be variable over that time with flux variability of $20 - 80$\%. \citet{Zapata2004a} tracked the ONC at 3.6 cm over four years, and identified 36 sources that are time variable by more than 30\%. More recently, \citet{Kounkel2014} mapped a large region of the Gould Belt at 4.5 GHz and 7.5 GHz over three epochs each separated by a month, and found 32 variable sources in the ONC. Futhermore, \citet{Rivilla2015} studied a field in the ONC at 0.7 cm and 0.9 cm and found 19 sources which are variable over long-term (monthly) timescales, and 5 sources which are variable on short timescales (hours to days). Moreover, very short timescale radio flares have been observed towards a number of pre-main sequence stars \citep[e.g.,][]{Bower2003,Forbrich2008,Rivilla2015}


Here we compare previously measured fluxes for detected compact radio sources with the fluxes in our maps. Time-baselines are $\sim$10 years at 1.3 cm and  3.6 cm and $\gtrsim$20 years at 6 cm, and we cannot characterize shorter timescales for variability. We thus seek to identify sources that may not have been detected as variable in previous, shorter time-baseline studies \citep{Felli1993b,Zapata2004a}. We also use our multiple epochs of 1.3 cm data to search for variability on timescales of $\sim7$ months, between November 10, 2013 to May 3, 2014.


As we discussed earlier, Sources 281, 391, 416, 423, 430, 433, 442, 512, 516, 537, 564, and 595 are very well resolved with the A-configuration at 1.3 cm. As such we exclude these sources from our variability considerations at 1.3 cm, as flux variations may be due to structure being resolved out rather than actual variability. 

We define a variable source as one for which the flux measurements are 3$\sigma$ discrepant from one epoch to the next, at any observed wavelength. $\Delta$F/F quantifies how variable a source is, where F is the mean flux of the source and $\Delta$F is the standard deviation of the fluxes. We show the results of this search in Table \ref{table:allvar}. For the sources detected in \citet{Zapata2004a} and \citet{Zapata2004b} we include a 10\% uncertainty on the flux on top of the uncertainties they quote to account for a systematic flux calibration uncertainty across the datasets.

At 1.3 cm we find 30 sources that show some indication of variability, with $\Delta$F/F ranging from 20-900\%. At 3.6 cm we identify 32 sources whose fluxes are variable, including 3 sources not identified as variable in \citet{Zapata2004a}, because they were too faint to be detected in individual epochs. The variability, as defined by $\Delta$F/F, of these sources ranges from 20-200\%. Finally, at 6 cm we identify 5 variable sources with $\Delta$F/F ranging from 50-100\%.

There were 13 sources detected by previous radio surveys of the ONC \citep[e.g.][]{Felli1993a,Felli1993b}. Most of those sources were not detected in the same bands as our observations, and so they are excluded from our variability analysis. However, the 5 variable sources with 6 cm fluxes from \citet{Felli1993b} were undetected in our maps, and have $\Delta$F/F ranging from 50-100\%. Given the high fluxes of the remainder of the sources, we would have expected to detect them in our maps, so those sources likely have similarly high variability amplitudes.

In all, we find that 55 of our sources are variable at one or more wavelengths. Of the variable sources, 11 are characterized as variable at multiple wavelengths. 20 are found to be variable at one wavelength but not another, although many of our constraints on $\Delta$F/F are not strong. The remaining sources could only be analyzed at a single wavelength.

We show the location of each variable source in the right panel of Figure \ref{fig:source_map}, with the strength of the variability ($\Delta$F/F) represented by the size of the plot symbol. We find that variability amplitude does not follow the same trend as free-free flux, with variability decreasing with increased separation from $\theta^1$ Ori C. Instead we find sources which are significantly variable out to large radii. Some of the most variable objects can be found at large separations.

Variability of radio emission from these sources is likely to arise from a few different mechanisms. It may be the result of gyrosynchrotron emission produced by magnetospheric activity in young stars \citep[e.g.,][]{Feigelson1999}. These flares may be the result of magnetic reconnections on the protostellar surface, which would produce radio flares on the timescales of minutes \citep[e.g][]{Dulk1985,Bower2003,Forbrich2008}. Interactions between the magnetic fields of the protostar and its disk could also produce flares on the timescales similar to the rotation periods of young stars, which are typically days to weeks in the ONC \citep[e.g.][]{Shu1997,Forbrich2006,Rodriguez2009}.

Free-free emission may also be variable if the density distribution of material being ionized is non-uniform causing the amount of ionized material to vary, or if the incident ionizing radiation is varying. Studies have found that O-type stars have winds that exhibit cyclical variability on timescales of hours to days \citep[see review by][]{Fullerton2003}. The visible, UV and X-ray intensity of $\theta^1$ Ori C varies with a period of 15.4 days \citep[e.g.,][]{Stahl1993,Stahl1996,Caillault1994,Walborn1994}, so the ionization level and therefore free-free flux might be expected to vary on a similar timescale. The ionized region, however, is likely to be many light days across or larger, so this variability may be washed out.

Inhomogeneities in the disk are unlikely to be brought into the ionized region on timescales shorter than the dynamical timescale. For disks, the dynamical timescale varies depending on location in the disk and the mass of the central star \citep[e.g.,][]{Kenyon2001}. Inner disk radii for young stars are found to be on the order of $0.1-1$ AU \citep[e.g.,][]{Eisner2007}, so the smallest dynamical timescales we can expect are on the order of weeks to half a year. Photoevaporation in disks tends to produce winds at radii larger than the critical radius, where the photoionized material has sufficient velocity to escape. For ionization by EUV photons this tends to occur at radii of $\gtrsim5$ AU \citep[e.g.,][]{Hollenbach1994,Gorti2007}, corresponding to dynamical timescales of a few years. Non-uniformities in the disk are therefore likely to cause longer term variability in the free-free emission.

Aside from the timescale of variability, the SED of the source at each epoch might be used to distinguish between free-free and synchrotron emission. As described in Section 3.2, free-free emission is characterized by a flat spectrum with $F_{\nu} \propto \nu^{-0.1}$. Gyrosynchrotron emission however, is expected to have a spectral index that is significantly negative, typically $F_{\nu} \propto \nu^{-0.7}$. We discuss constraints on the nature of some of these sources in Section 4.3. Further studies with concurrent flux measurements at multiple wavelengths, however, are needed to fully distinguish between these sources of emission.

Here we do not have simultaneous flux measurements at all bands for each epoch of data, so it is difficult to constrain the spectral index of the emission at each epoch. There are, however, a few sources which change flux significantly between the 1.3 cm observations on March 3 and March 7 2014. For example, on March 3, Source 529 had a 1.3 cm flux of $7$ mJy, but on March 7 it was down to a flux of 4.5 mJy. By May 7th the flux was all the way down at 0.5 mJy. Such an extreme change in flux may be indicative of gyrosynchrotron emission from a magnetic flare. Source 544 also shows a similar pattern. Source 587 has a flux of 1.7 mJy on March 3, but on March 7 it's flux increased significantly to 22 mJy, again possibly indicative of a magnetic flare. For most sources, however, we do not have sufficient time resolution to distinguish between daily, weekly, or even longer variability timescales.

\subsection{Nature of Detected Sources}

We detected emission in at least one of our maps from 67 {\it HST} identified proplyds \citep{Ricci2008,Mann2010,Mann2014}. Furthermore, we have detected radio emission towards 120 sources that have been identified by near-infrared imaging \citep{Hillenbrand2000}. We also detect radio emission from 2 sources dubbed `MM' by \citet{Eisner2008}, indicating that they have previously only been detected at wavelengths longer than 1 mm. Finally, we have detected 51 sources that are not associated with a known proplyd or near-infrared detected source.

The majority of our targets, including all of the sources identified as proplyds, are well fit by our free-free and dust emission model, in agreement with previous conclusions that these objects are disks with winds driven by photoevaporation \citep[e.g.,][]{Churchwell1987,Henney1998}. We detect a turnover in the free-free emission spectrum for 40 objects, as evidenced by $5.5 \text{GHz} < \nu_{turn} < 22 \text{GHz}$. There are at least 3 sources (Sources 374, 465, 473) that might even have $\nu_{turn} > 22\text{GHz}$, indicating that our maps are insufficient to fully characterize their emission. With such high turnover frequencies, these objects must have small inner boundaries to their ionized envelopes and are likely very compact and dense. Further short wavelength observations are necessary to better constrain the free-free emission spectrum.

Some sources have SEDs that appear to be fitted well by our free-free + dust model with some variability included. Fluxes at all three wavelengths were measured concurrently on March 3, 2014, and if we just consider those flux measurements, all of our sources are fitted well by free-free emission models. Without simultaneous 3.6 cm and 6 cm measurements for the other 1.3 cm epochs it is impossible to say whether the SEDs at those epochs remain consistent with free-free emission, although it seems probable.

Below we split the sources whose SEDs are not fitted well by our model and therefore are not indicative of being free-free emission, or do not follow the expected trend of decreasing centimeter flux with increasing separation from $\theta^1$ Ori C:

\subsubsection{Strong Free-Free Sources}

{\bf Source 418} is $\theta^1$ Ori A, a binary system with a B0.5 primary star and a low-mass companion, possibly a T Tauri star \citep{Levato1976,Bossi1989}, which is known to be highly variable \citep[e.g.,][]{Felli1993b}. \citet{Rivilla2015} suggest that this variability may be twofold, (i) variations in free-free opacity from a stellar wind from the interactions with the companion, and (ii) variations in the non-thermal emission from stellar activity related to the distance between to binary, similar to the case of WR 140 \citep[e.g.,][]{Williams1990}. \citet{Rivilla2015} suggest that while the former mechanism may be present, the latter is required to explain previous observations.

{\bf Sources 279 and 308} each have radio spectra that are steeper than $\nu^{0.6}$. Source 279 is the Becklin-Neugebauer Object, and is thought to be a runaway B star, ejected from a system with Source I (our Source 308) in an explosive event 500 years ago \citep[e.g.,][]{Plambeck1995,Gomez2008,Plambeck2013}. The BN Object has a spectral dependence of $\nu^{1.3}$ below 100 GHz, above which it flattens, and is suggested to be free-free emission from a dense, hypercompact HII region. Source I has a spectral dependence of $\nu^2$ and is most easily explained by H$^-$ free-free emission in a disk \citep[e.g.,][]{Plambeck2013}. Both sources have massive stars driving ionizing circumstellar material and driving the free free emission we detect, explaining their significant fluxes despite their distance from $\theta^1$ Ori C.

\subsubsection{Dust-Only Sources}

{\bf Sources 134, 236, 246, and 301} have millimeter (850 $\mu$m or 1.3mm) detections and are detected at 1.3 cm in our maps, but are undetected at 3.6 cm and 6 cm. The SEDs for all of these sources are well fit by a model that is predominantly dust emission at 1.3 cm (and perhaps a minor contribution from free-free emission). Sources 236 and 246 are identified by \citet{Eisner2008} as ``MM" objects (MM21 and MM8 respectively), which lack near-IR counterparts. Source 301 is also identified by \citet{Eisner2008} as LMLA 162. All three of those sources (236, 246, and 301) are among the most massive known sources in the ONC \citep[$>0.2$ $M_{\odot}$][]{Eisner2008}. They are likely highly embedded young objects, and may be candidate Class 0 or I objects as suggested by \citet{Eisner2008}.

\subsubsection{Non-Thermal Radio Sources}

{\bf Sources 11 and 617} each have spectra with steep negative spectral indices (see Figure \ref{fig:radial_fluxes}), which may indicate that they are emitting synchrotron radiation. Source 11 is not associated with any previous detections in our reference catalogs, and is not found to not be variable at 3.6 cm. Source 617 has previously been detected, and is commonly referred to as F \citep[e.g.,][]{Churchwell1987,Garay1987,Felli1993a}. It has previously been found to experience radio flares on timescales as short as hours and possibly as long as months \citep{Rivilla2015}.

{\bf Sources 154, 394, 437, 440, 529, 544, 587, and 730} are highly variable sources, showing significant changes in flux over just a few days between our observations on March 3, 2014 and March 7, 2014. All are significant outliers in Figure \ref{fig:radial_fluxes}. Source 154 has previously been identified as A \citep[e.g.,][]{Churchwell1987,Garay1987,Felli1993a}, and \citet{Zapata2004a} find the source to show large percentages of circular polarization, and suggest that the emission may be gyrosynchrotron in nature. \citet{Felli1993b} also classify Source 587, also known as G, as a non-thermal variable emitter. All of these sources are associated with infrared detected objects \citep{Hillenbrand2000}. These sources may be indicative of radio flares of gyrosynchrotron emission, but further observations with concurrent flux measurements at multiple wavelengths are needed to confirm this.

{\bf Source 903} is located far from $\theta^1$ Ori C, and is only detected at 6 cm, but it has a high flux given it's significant separation (see Figure \ref{fig:radial_fluxes}). It is out of the field of view of our 1.3 cm data, and right on the edge of our 3.6 cm map, but undetected. It is associated with the proplyd 281-306, which is a disk seen only in silhouette with HST \citep{Ricci2008}. Radio emission from this source may be attributed to magnetic activity from the young star or free-free emission from material ionized by the star itself, as it is likely too far to be material ionized by $\theta^1$ Ori C.

{\bf Source 904} is also located far from $\theta^1$ Ori C, with a high flux given it's separation (Figure \ref{fig:radial_fluxes}), and is outside the field of view of our 1.3 cm map. It's 6 cm and 3.6 cm fluxes are consistent with free-free emission, but do show indications that the spectral index may be significantly negative. It is unassociated with any previous catalog.

\subsubsection{Extragalactic Sources}


Given the large survey area of our maps, it is possible that some of our detections are extragalactic in nature. Following \citet{Fomalont1991}, at 6 cm we would expect the number of extragalactic contaminants greater than 156 $\mu$Jy (our $6\sigma$ threshold) in our 109 arcmin$^2$ survey area to be $6.5\pm2.3$. At 3.6 cm, using \citet{Fomalont2002}, we estimate $1.1\pm0.2$ extragalactic sources in our 70 arcmin$^2$ survey area to above 216 $\mu$Jy. No similar survey exists at 1.3 cm, so we use the 3.6 cm numbers to estimate that in our 1.3 cm map we would expect $1.8\pm0.2$ contaminants in our 34 arcmin$^2$ survey area above 72 $\mu$Jy. As most of these sources show non-thermal emission with negative spectral indices at these wavelengths \citep{Condon1992} they should be fainter at 1.3 cm than at 3.6 cm and therefore we would expect the contamination at 1.3 cm to be even smaller than this.

\subsection{Free-free Contamination of Sub-millimeter Dust Masses}

At submillimeter, millimeter and radio wavelengths, the light emitted by dust grains is expected to be largely optically thin and the flux is directly proportional to the amount of dust present \citep[e.g.,][]{Beckwith1990}. As such, submillimeter flux measurements of protoplanetary disks are commonly used to measure the mass of those disks \citep[e.g.,][]{Andrews2005,Eisner2008,Mann2010,Andrews2013,Mann2014}. A number of previous surveys across millimeter and submillimeter wavelengths have employed this method to measure disk masses for protoplanetary disks in the ONC \citep[e.g.,][]{Mundy1995,Bally1998b,Williams2005,Eisner2006,Eisner2008,Mann2010,Mann2014}.

The proplyds, however, are located near the Trapezium cluster of young massive stars that are photoevaporating the disks \citep[e.g.,][]{Churchwell1987}. The ionized material produced in the proplyds emits free-free emission, which can be bright at the same wavelengths used to measure disk masses. In order to accurately measure disk masses, it is therefore important to separate the dust and free-free contributions to sub-millimeter and millimeter fluxes. This is particularly true with the advent of ALMA, which will detect disks in the ONC much fainter than those that have been previously detected.

In this work we characterized the free-free emission from a collection of compact radio sources in the ONC. In Table \ref{table:modelparams} we use the best fit free-free emission spectrum model from Section 3.2 to calculate the expected free-free flux at all ALMA bands. This table can be used to correct measured millemeter fluxes for free-free contamination, and accurately measure the sub-millimeter dust flux and thereby the dust mass.

For most sources this extrapolation provides a good estimate of the free-free contribution of the source at ALMA wavelengths. This is not true, however, of sources for which the model fit is poor as was discussed in the previous section. Furthermore, the extrapolation to ALMA bands for sources whose turnover frequency is designated as $>22$ GHz is also very uncertain. Many of these sources were only detected at 1.3 cm, and a few have radio photometry that is best fit by a $\nu^{0.6}$ power law. Our extrapolation for these sources assumes $\nu_{turn} = 22$ GHz, but if $\nu_{turn} > 22$ GHz the free-free flux at ALMA bands would be greater than our current prediction. Further radio or millimeter observations are necessary to constrain $\nu_{turn}$ before accurate ALMA free-free fluxes can be predicted.

Variability is also a significant source of uncertainty in determining how well free-free emission from disks can be constrained and removed from disk mass measurements, if the free-free flux to dust flux ratio is large. For example, the measured 230 GHz flux of Source 439 is 8.8 mJy and the model free-free flux at 230 GHz is 2.8 mJy, with a variability of 24\%. So free-free emission makes up 32$\pm$ 8\% of the total 230 GHz flux. Source 466, however, has a measured 230 GHz flux of 7.7 mJy, 0.3 mJy of which is due to free free emission with 50.8\% variability, so the free-free emission makes up  5$\pm$ 2\% of that flux. Because of the smaller free-free flux to total flux ratio of Source 466, the dust flux can be better constrained, even though Source 466 is more variable. 

ALMA, however, will be able to detect disks which are much fainter than has previously been possible, For these sources, variability may be a significant problem. Source 469 has a 230 GHz free-free flux of 0.33 mJy with a variability of 108\%. Although it has no previous millimeter detections, if it were found to have a millimeter flux of 0.5 mJy, the dust mass calculation would be highly uncertain because the free-free flux would make up 65$\pm$70\% of the total 230 GHz flux.

An accurate estimate of the uncertainty associated with variability of our sources, however, likely requires further monitoring of the sources to characterize the timescale and amplitude of the variability. Sources with significant variability may even require concurrent millimeter and radio flux measurements in order to measure the free-free contribution to millimeter flux measurements.

While not important for some objects, the correction for free-free emission is often crucial for correctly measuring disk mass. For example, the free-free emission from sources 391, 408, 416, 418, 421, 423, 430, 438, 439, 442, 446, 460, 465, 484, 491, 494, 499, 512, 516, 518, 535, 537, 555, 564, 605, 612 and 617 contributes $>$50\% of the measured 3 mm fluxes. Free-free emission also contributes $>$40\% of the measured 1.3 mm fluxes for sources 408, 416, 418, 421, 423, 438, 442, 460, 465, 484, 491, 494, 499, 516, 518, 535, 564 and 617 and $>$30\% of the measured 850 $\mu$m fluxes for sources 408, 421, 423, 438, 460, 465, 484, 491, 494, 499 and 617. Without these corrections, disk mass estimates from these sub-millimeter bands would be off by a significant amount.

\subsection{Future Work}

While our radio dataset is tremendously useful for finding radio sources and characterizing their free-free emission for ALMA disk mass studies in the ONC, the data also has a number of other applications which we will explore in future work.

Due to the high resolution of our maps, particularly at 1.3 cm, many of the sources detected in our maps are well resolved. The morphologies of these objects show interesting features, particularly when matched up with high resolution HST images of the proplyds. For many sources structure in HST maps are well matched with features in our maps.

Furthermore, we can use resolved images to measure mass loss rates for the protoplanetary disks. The free-free emission we detect here originates from ionized cocoons of gas which are flowing away from their associated disks under the intense radiation pressure from the star $\theta^1$ Ori C. The flux of this free free emission coupled with measured sizes of these cocoons of gas is sufficient to measure disk mass-loss rates. Mass loss rates have previously been measured for a handful of disks in the ONC \citep[e.g.,][]{Churchwell1987}, but the improved sensitivity and resolution of our maps will allow us to make this measurement for many more sources.

\section{Conclusions}

We have produced new high spatial resolution maps of the Orion Nebula at 1.3 cm, 3.6 cm and 6 cm with significantly improved sensitivities compared with previous radio studies of the region using the JVLA. In these maps we search for compact ($\lesssim$2") radio sources, and use these detections to constrain the properties of free-free emission from protoplanetary disks in the ONC. Free-free emission is emitted from the ionized winds driven by the nearby massive star $\theta^{1}$ Ori C. Constraints on this free-free emission are crucial for studies aiming to measure disk masses for the proplyds from sub-millimeter fluxes.

We detect 144 sources at 1.3 cm, 98 sources at 3.6 cm, and 108 sources at 6 cm, for a total of 175 unique sources. Of these 175 detections, 149 have previously been detected at radio wavelengths, 67 are associated with {\it HST} detected proplyds, 120 with near-infrared detected YSOs, 40 with sources detected previously at millimeter wavelengths, and 11 are detected for the first time at any wavelength.

For each source detected in our maps we report its position and flux, as measured by fitting a gaussian to the source, in Table \ref{table:fluxes}. For previously identified sources not detected in one or more of our maps we also report the integrated flux in an 1" aperture measured towards the source. This information is presented in an extended version of Table \ref{table:fluxes} that is available in the online materials.

We fit each of our source spectra with a combined dust + free-free emission model. The majority of our targets are fitted well by this dust + free-free model, with many showing evidence for a turnover in the free-free emission. Further studies of free-free emission may benefit from longer wavelength flux measurements to better constrain the free-free turnover. Four of our detected sources (134, 236, 246, and 301) have SEDs that are consistent with being produced entirely by dust emission and are likely highly embedded young objects. We also detect the Becklin-Neugebauer Object, it's alleged counterpart Source I, and $\theta^1$ Ori A.

Many of our sources have previously measured radio fluxes, so we can investigate variability. We find that 30 sources are variable at 1.3 cm, 32 at 3.6 cm, and 5 at 6 cm. 55 of our detected sources are variable at one or more wavelengths. For sources that are variable we define a metric, $\Delta$F/F, to quantify the variability, and find that $\Delta$F/F $\approx 20-900$\% for our targets. 13 are variable at $>100$\%, suggesting that any sub-millimeter measurements will be very uncertain. The time sampling is, however, poor, so more dedicated monitoring of our targets is necessary for better understanding this variability.

Finally, the free-free emission properties derived from our modeling can be extrapolated to sub-millimeter wavelengths to estimate the free-free contribution to sub-millimeter fluxes. This is necessary for correctly distinguishing dust emission and free-free emission at sub-millimeter wavelengths, particularly when sub-millimeter fluxes are used to calculate disk dust masses. This will be crucial for future studies of dust emission from protoplanetary disks in the ONC with ALMA. We provide free-free flux estimates for each detected source at each ALMA band in Table \ref{table:modelparams}. Variability is a significant source of uncertainty in correcting millimeter flux measurements for free-free emission if the free-free flux to dust flux ratio is large, so understanding this variability is an important future direction.

In the future we will use this dataset to study the morphologies of the sources resolved in our high resolution VLA maps, particularly as compared with HST images of the proplyds. We will also measure the rate at which material is being photoevaporated and lost from the disks of these sources under the intense radiation and winds from $\theta^1$ Ori C and the Trapezium Cluster, and therefore derive disk lifetimes for the protoplanetary disks in the ONC.

\acknowledgements

This material is based upon work supported by the National Science Foundation Graduate Research Fellowship under Grant No. 2012115762.
This work was supported by NSF AAG grant 1311910.
The results reported herein benefitted from collaborations and/or information exchange within NASA's Nexus for Exoplanet System Science (NExSS) research coordination network sponsored by NASA's Science Mission Directorate.
The National Radio Astronomy Observatory is a facility of the National Science Foundation operated under cooperative agreement by Associated Universities, Inc.

\bibliographystyle{apj}
\bibliography{ms}

\tabletypesize{\tiny}
\LongTables
\clearpage
\begin{landscape}
\begin{turnpage}
\global\pdfpageattr\expandafter{\the\pdfpageattr/Rotate 90}
\begin{deluxetable*}{ccccccccrrrrrrr}
\tablehead{\colhead{ID} & \colhead{Proplyd Name} & \colhead{HC00 ID} & \colhead{GMR ID} & \colhead{Z04a ID} & \colhead{Other Names} & \colhead{R.A.} & \colhead{Dec} & \colhead{$F_{\nu,6\text{cm}}$} & \colhead{$F_{\nu,3.6\text{cm}}$} & \colhead{$F_{\nu,1.3\text{cm},1}$\tablenotemark{a}} & \colhead{$F_{\nu,1.3\text{cm},2}$\tablenotemark{a}} & \colhead{$F_{\nu,1.3\text{cm},3}$\tablenotemark{a}} & \colhead{$F_{\nu,1.3\text{cm},4}$\tablenotemark{a}} & \colhead{$F_{\nu,1.3\text{cm},mean}$\tablenotemark{a}}\\ \colhead{ } & \colhead{ } & \colhead{ } & \colhead{ } & \colhead{ } & \colhead{ } & \colhead{[J2000]} & \colhead{[J2000]} & \colhead{[mJy]} & \colhead{[mJy]} & \colhead{[mJy]} & \colhead{[mJy]} & \colhead{[mJy]} & \colhead{[mJy]} & \colhead{[mJy]}}
\startdata
1 & 4538-311 & -- & -- & -- & -- & 5h34m53.79s & -5d23m10.73s & $ 0.015 \pm  0.567$ & \nodata & \nodata & \nodata & \nodata & \nodata & \nodata \\
2 & -- & -- & -- & -- & -- & 5h34m55.97s & -5d23m13.00s & $ 0.269 \pm  0.338$ & \nodata & \nodata & \nodata & \nodata & \nodata & \nodata \\
3 & 4582-635 & -- & -- & -- & -- & 5h34m58.16s & -5d26m35.13s & $-0.245 \pm  0.409$ & \nodata & \nodata & \nodata & \nodata & \nodata & \nodata \\
4 & 005-514 & -- & -- & -- & -- & 5h35m00.47s & -5d25m14.34s & $ 0.114 \pm  0.233$ & \nodata & \nodata & \nodata & \nodata & \nodata & \nodata \\
5 & 006-439 & -- & -- & -- & -- & 5h35m00.58s & -5d24m38.79s & $-0.008 \pm  0.229$ & $-0.049 \pm  1.203$ & \nodata & \nodata & \nodata & \nodata & \nodata \\
6 & 016-149 & -- & -- & -- & -- & 5h35m01.60s & -5d21m49.35s & $-0.262 \pm  0.274$ & $-0.149 \pm  0.884$ & \nodata & \nodata & \nodata & \nodata & \nodata \\
7 & 038-627 & -- & -- & -- & -- & 5h35m04.19s & -5d26m27.89s & $-0.061 \pm  0.237$ & \nodata & \nodata & \nodata & \nodata & \nodata & \nodata \\
8 & 044-527 & -- & -- & -- & -- & 5h35m04.42s & -5d25m27.40s & $ 0.006 \pm  0.174$ & $-0.091 \pm  0.609$ & \nodata & \nodata & \nodata & \nodata & \nodata \\
\enddata
\label{table:fluxes}
\tablenotetext{a}{Subscript 1 indicates the 1.3 cm data were taken on Nov 10, 2013, 2 on Mar. 3, 2014, 3 on Mar. 7, 2014, 4 on May 3, 2014, and mean indicates the image from the combined 1.3 cm tracks.}
\end{deluxetable*}

\end{turnpage}
\clearpage
\end{landscape}

\LongTables
\clearpage
\begin{landscape}
\begin{deluxetable*}{crrrrrrrrrrrrr}
\tablehead{\colhead{ID} & \colhead{$\nu_{\text{turn}}$} & \colhead{$F_{\nu,\text{turn}}$} & \colhead{$F_{\nu,\text{dust},230\text{GHz}}$} & \colhead{$F_{\nu,\text{Band 1}}$\tablenotemark{a}} & \colhead{$F_{\nu,\text{Band 2}}$\tablenotemark{a}} & \colhead{$F_{\nu,\text{Band 3}}$\tablenotemark{a}} & \colhead{$F_{\nu,\text{Band 4}}$\tablenotemark{a}} & \colhead{$F_{\nu,\text{Band 5}}$\tablenotemark{a}} & \colhead{$F_{\nu,\text{Band 6}}$\tablenotemark{a}} & \colhead{$F_{\nu,\text{Band 7}}$\tablenotemark{a}} & \colhead{$F_{\nu,\text{Band 8}}$\tablenotemark{a}} & \colhead{$F_{\nu,\text{Band 9}}$\tablenotemark{a}} & \colhead{$F_{\nu,\text{Band 10}}$\tablenotemark{a}}\\ \colhead{ } & \colhead{[GHz]} & \colhead{[mJy]} & \colhead{[mJy]} & \colhead{[mJy]} & \colhead{[mJy]} & \colhead{[mJy]} & \colhead{[mJy]} & \colhead{[mJy]} & \colhead{[mJy]} & \colhead{[mJy]} & \colhead{[mJy]} & \colhead{[mJy]} & \colhead{[mJy]}}
\startdata
11 & $<5.5$ & $ 3.188 \pm  0.161$ & \nodata & 2.65 & 2.44 & 2.39 & 2.29 & 2.23 & 2.19 & 2.11 & 2.05 & 1.98 & 1.93 \\
18 & $<5.5$ & $ 0.582 \pm  0.078$ & \nodata & 0.48 & 0.45 & 0.44 & 0.42 & 0.41 & 0.40 & 0.38 & 0.37 & 0.36 & 0.35 \\
69 & $>22.0$ & $ 0.265 \pm  0.181$ & \nodata & 0.25 & 0.23 & 0.23 & 0.22 & 0.21 & 0.21 & 0.20 & 0.20 & 0.19 & 0.18 \\
76 & $<5.5$ & $ 0.237 \pm  0.021$ & \nodata & 0.20 & 0.18 & 0.18 & 0.17 & 0.17 & 0.16 & 0.16 & 0.15 & 0.15 & 0.14 \\
77 & $11.477 \pm  6.002$ & $ 0.382 \pm  0.055$ & \nodata & 0.34 & 0.31 & 0.31 & 0.30 & 0.29 & 0.28 & 0.27 & 0.26 & 0.25 & 0.25 \\
97 & $16.116 \pm  8.250$ & $ 0.182 \pm  0.040$ & \nodata & 0.17 & 0.16 & 0.15 & 0.15 & 0.14 & 0.14 & 0.13 & 0.13 & 0.13 & 0.12 \\
103 & $>22.0$ & $ 0.165 \pm  0.048$ & \nodata & 0.16 & 0.15 & 0.14 & 0.14 & 0.13 & 0.13 & 0.13 & 0.12 & 0.12 & 0.11 \\
109 & $ 9.152 \pm  8.250$ & $ 0.057 \pm  0.090$ & \nodata & 0.05 & 0.05 & 0.04 & 0.04 & 0.04 & 0.04 & 0.04 & 0.04 & 0.04 & 0.04 \\
\enddata
\label{table:modelparams}
\tablenotetext{a}{Band 1 = 35 GHz, Band 2 = 80 GHz, Band 3 = 100 GHz, Band 4 = 150 GHz, Band 5 = 200 GHz, Band 6 = 230 GHz, Band 7 = 345 GHz, Band 8 = 450 GHz, Band 9 = 650 GHz, Band 10 = 850 GHz}
\end{deluxetable*}

\clearpage
\end{landscape}

\global\pdfpageattr\expandafter{\the\pdfpageattr/Rotate 0}
\tabletypesize{\small}
\LongTables
\begin{deluxetable*}{ccrcrcr}
\tablehead{\colhead{ID} & \colhead{1.3 cm Variable?} & \colhead{$(\Delta$F/F$)_{1.3\text{cm}}$} & \colhead{3.6 cm Variable?} & \colhead{$(\Delta$F/F$)_{3.6\text{cm}}$} & \colhead{6 cm Variable?} & \colhead{$(\Delta$F/F$)_{6\text{cm}}$}}
\startdata
11 & N & \nodata & Y & 20 $\pm$  28 & -- & \nodata \\
76 & N & \nodata & -- & \nodata & -- & \nodata \\
77 & N & \nodata & Y & 144 $\pm$  25 & -- & \nodata \\
102 & -- & \nodata & N & \nodata & -- & \nodata \\
103 & N & \nodata & -- & \nodata & -- & \nodata \\
115 & Y & 78 $\pm$ 102 & N & \nodata & -- & \nodata \\
134 & N & \nodata & -- & \nodata & -- & \nodata \\
151 & N & \nodata & N & \nodata & -- & \nodata \\
\enddata
\label{table:allvar}
\end{deluxetable*}

\begin{figure*}
\centering
\includegraphics[width=4in]{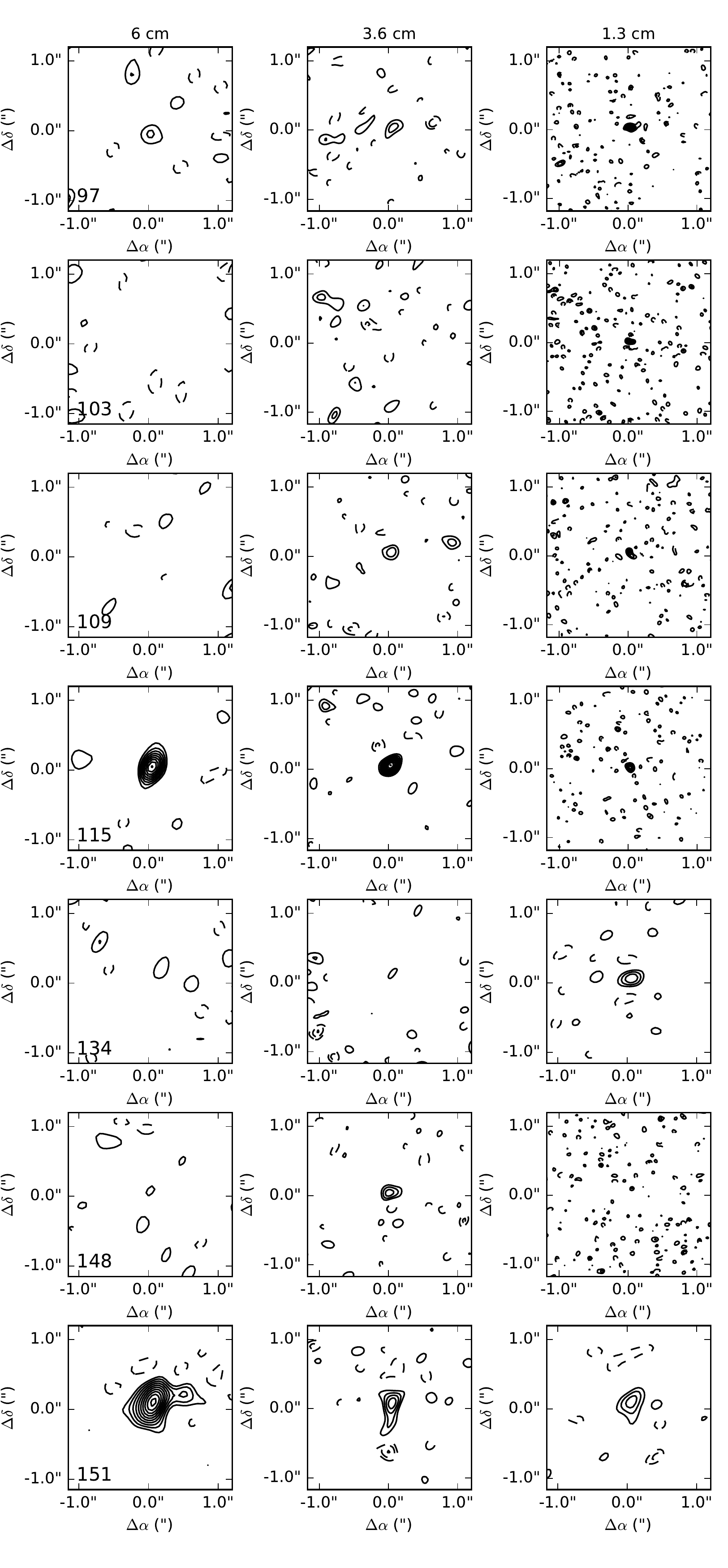}
\caption{{\it Continued}}
\end{figure*}

\begin{figure*}
\centering
\includegraphics[width=4in]{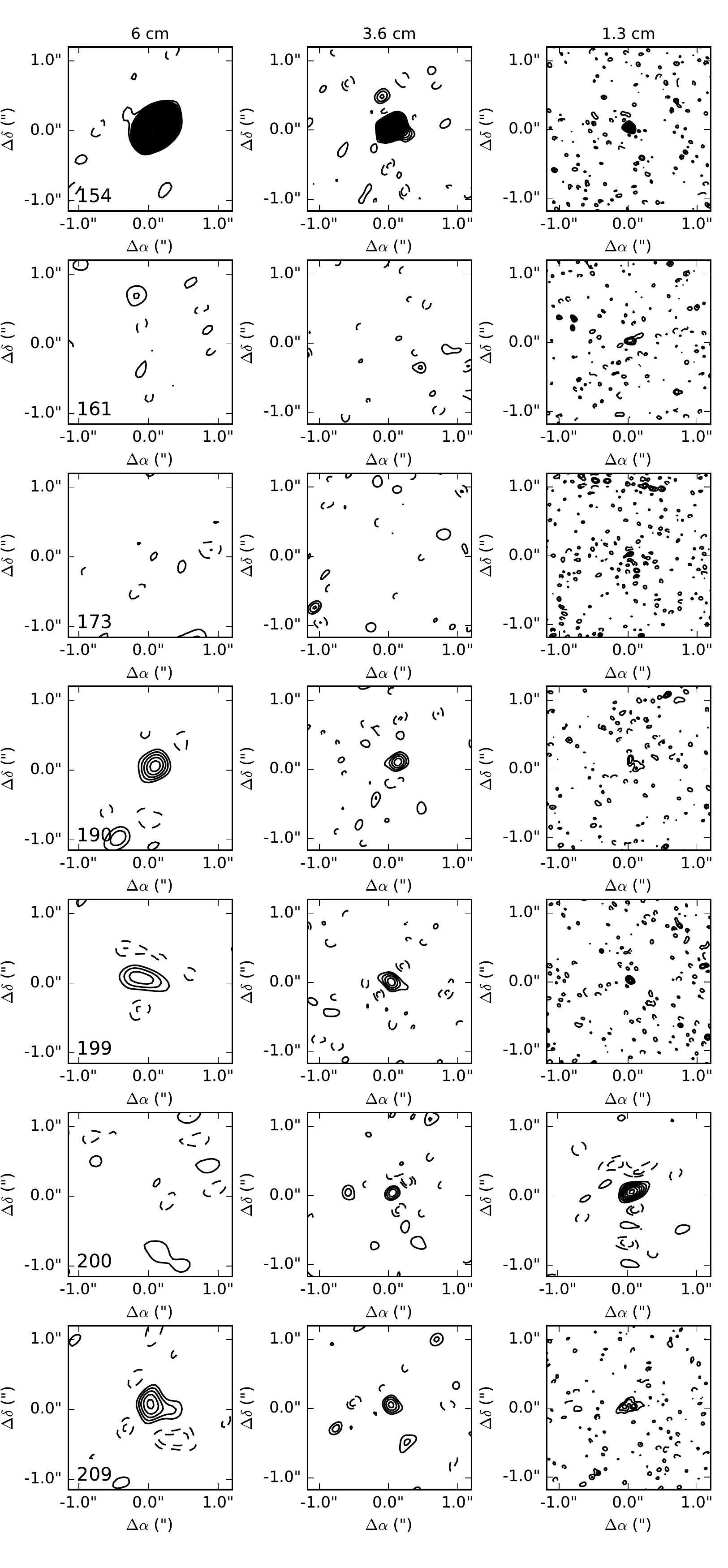}
\caption{{\it Continued}}
\end{figure*}

\begin{figure*}
\centering
\includegraphics[width=4in]{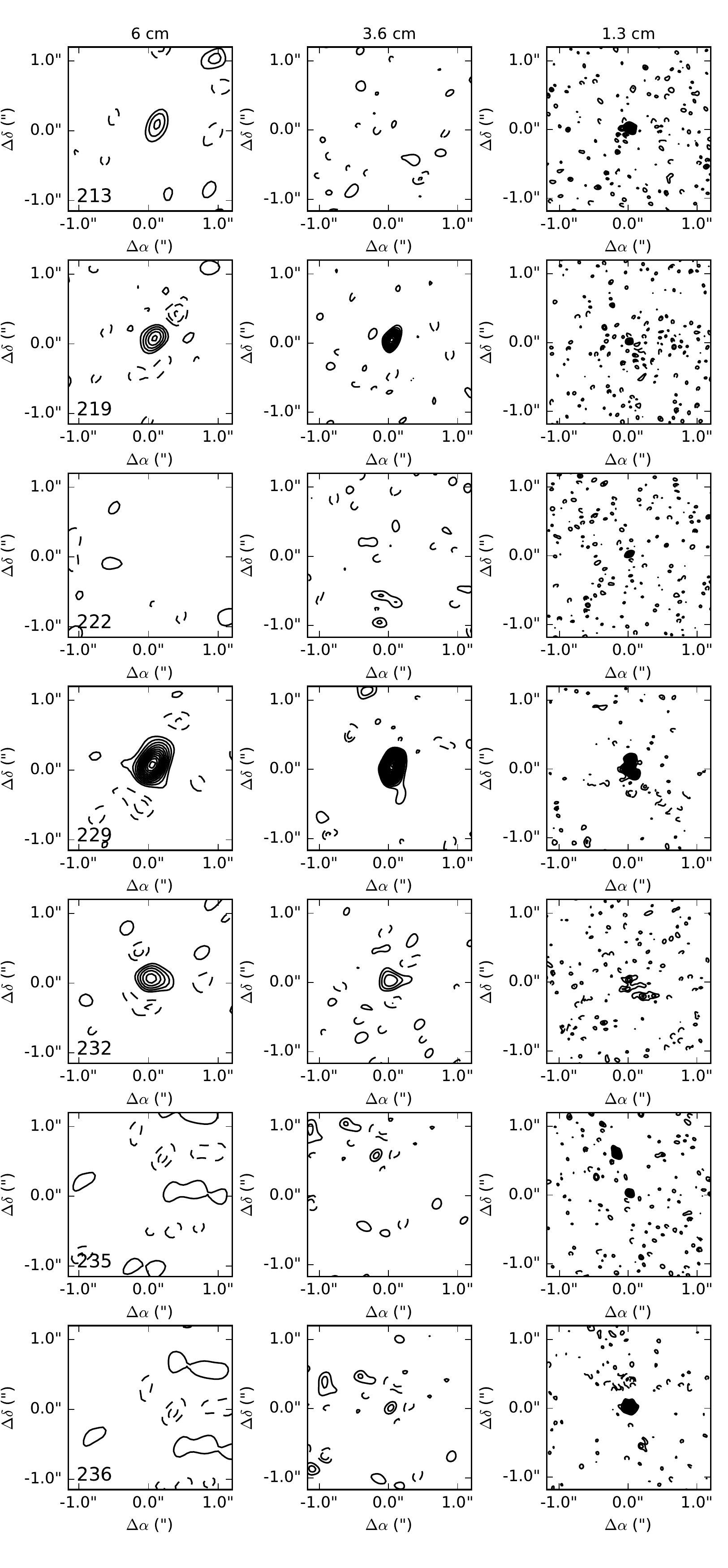}
\caption{{\it Continued}}
\end{figure*}

\begin{figure*}
\centering
\includegraphics[width=4in]{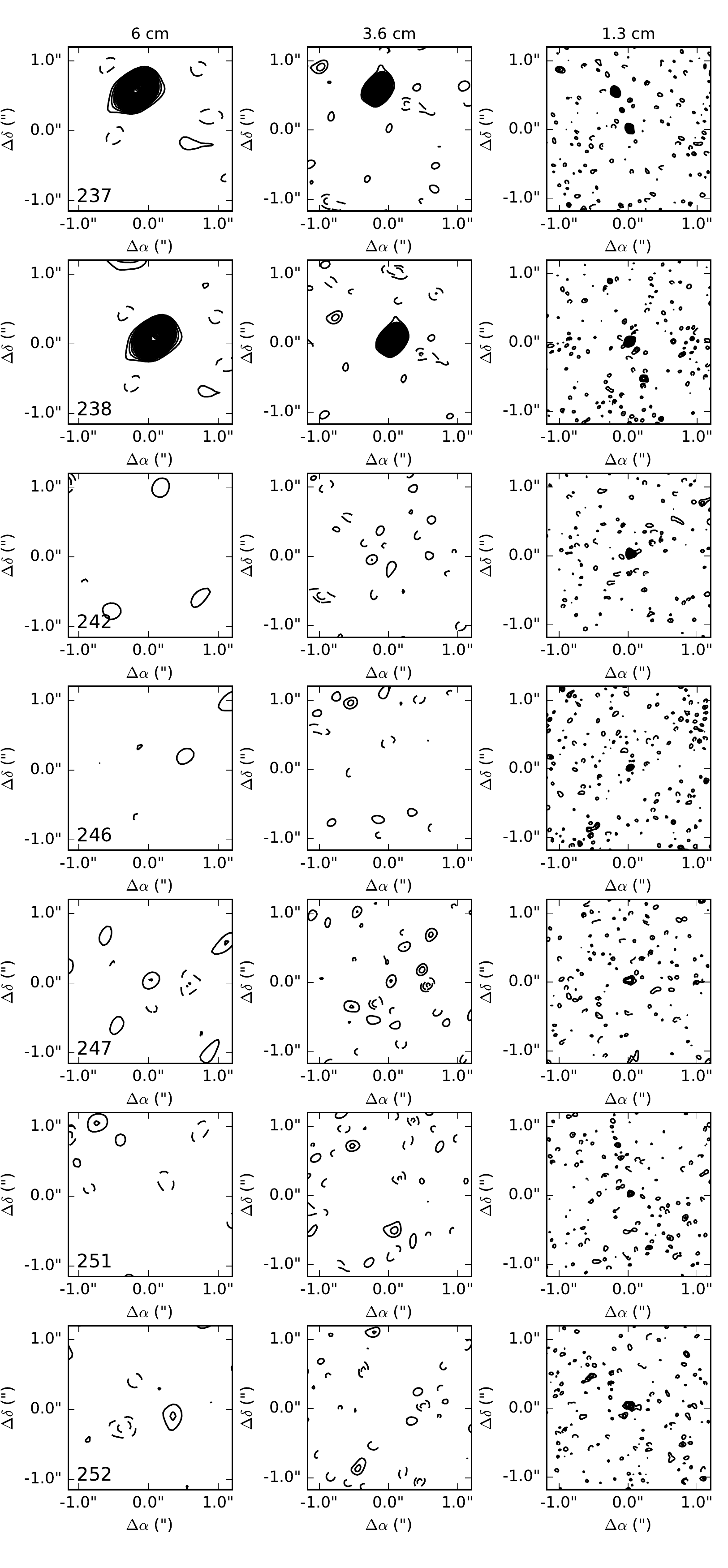}
\caption{{\it Continued}}
\end{figure*}

\begin{figure*}
\centering
\includegraphics[width=4in]{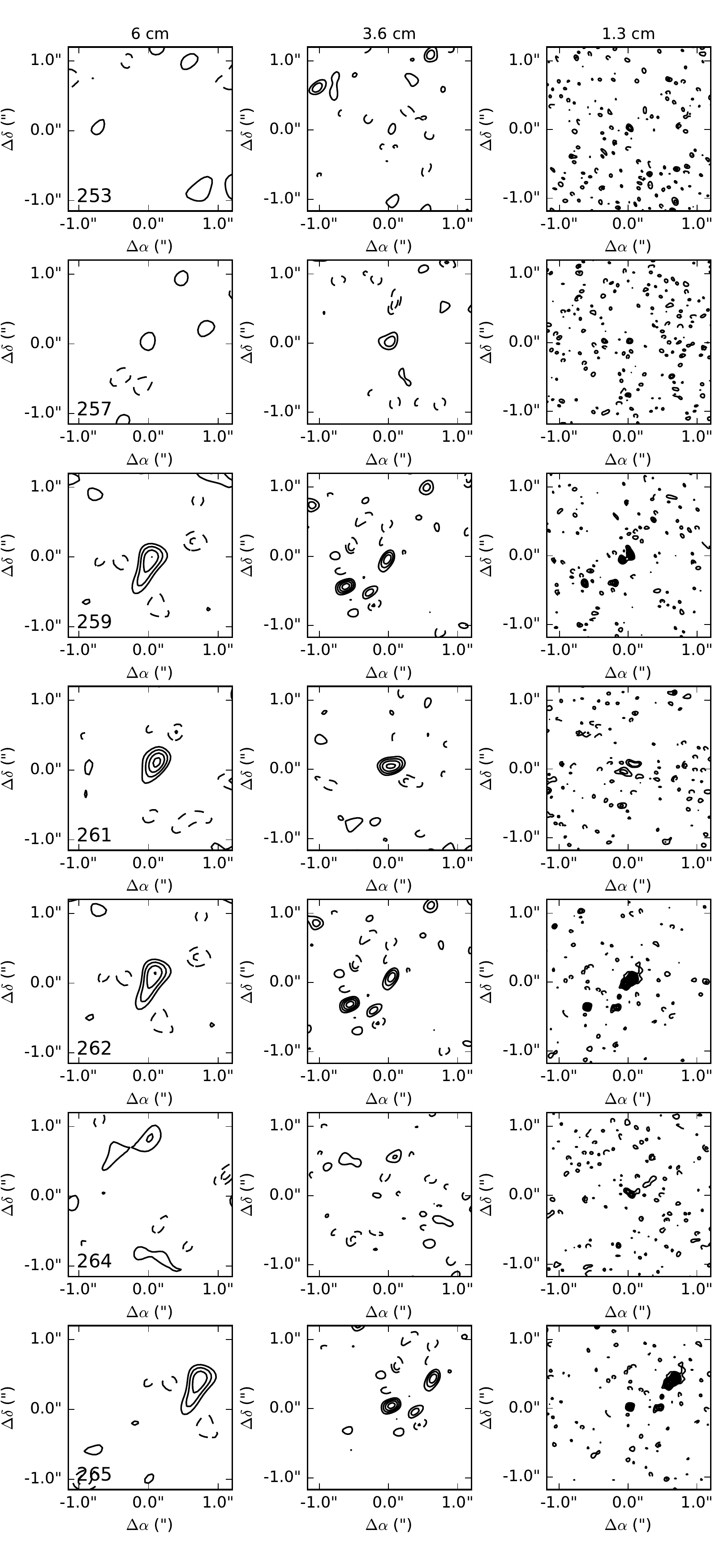}
\caption{{\it Continued}}
\end{figure*}

\begin{figure*}
\centering
\includegraphics[width=4in]{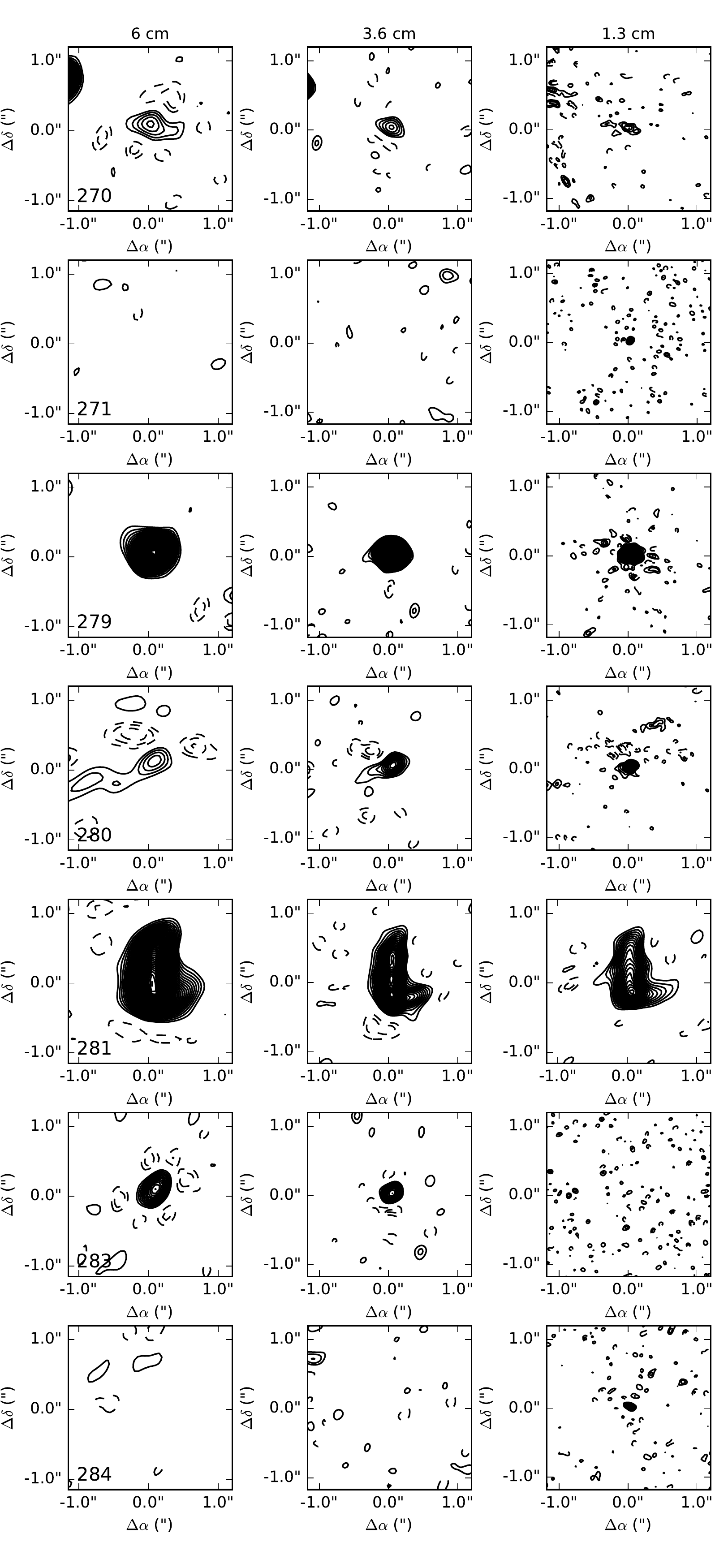}
\caption{{\it Continued}}
\end{figure*}

\begin{figure*}
\centering
\includegraphics[width=4in]{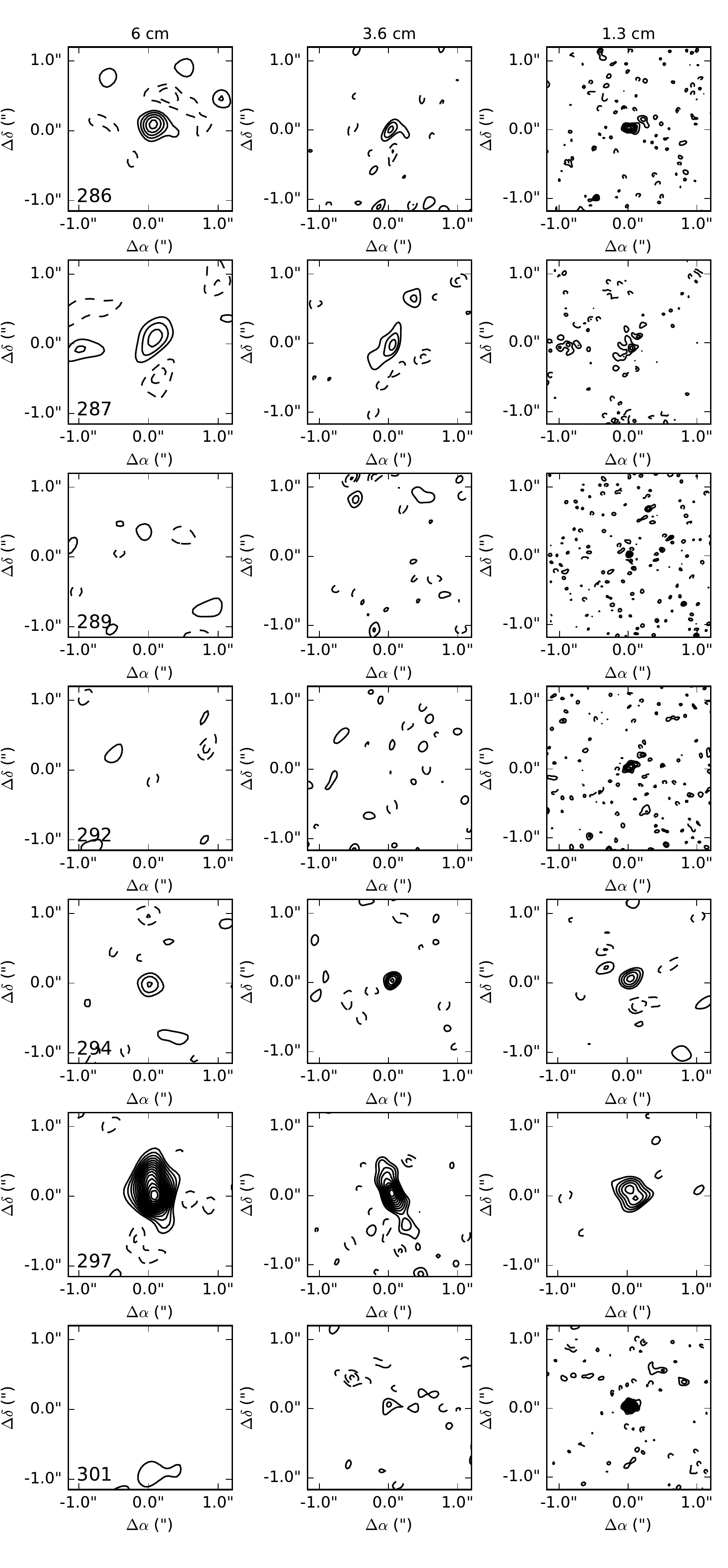}
\caption{{\it Continued}}
\end{figure*}

\begin{figure*}
\centering
\includegraphics[width=4in]{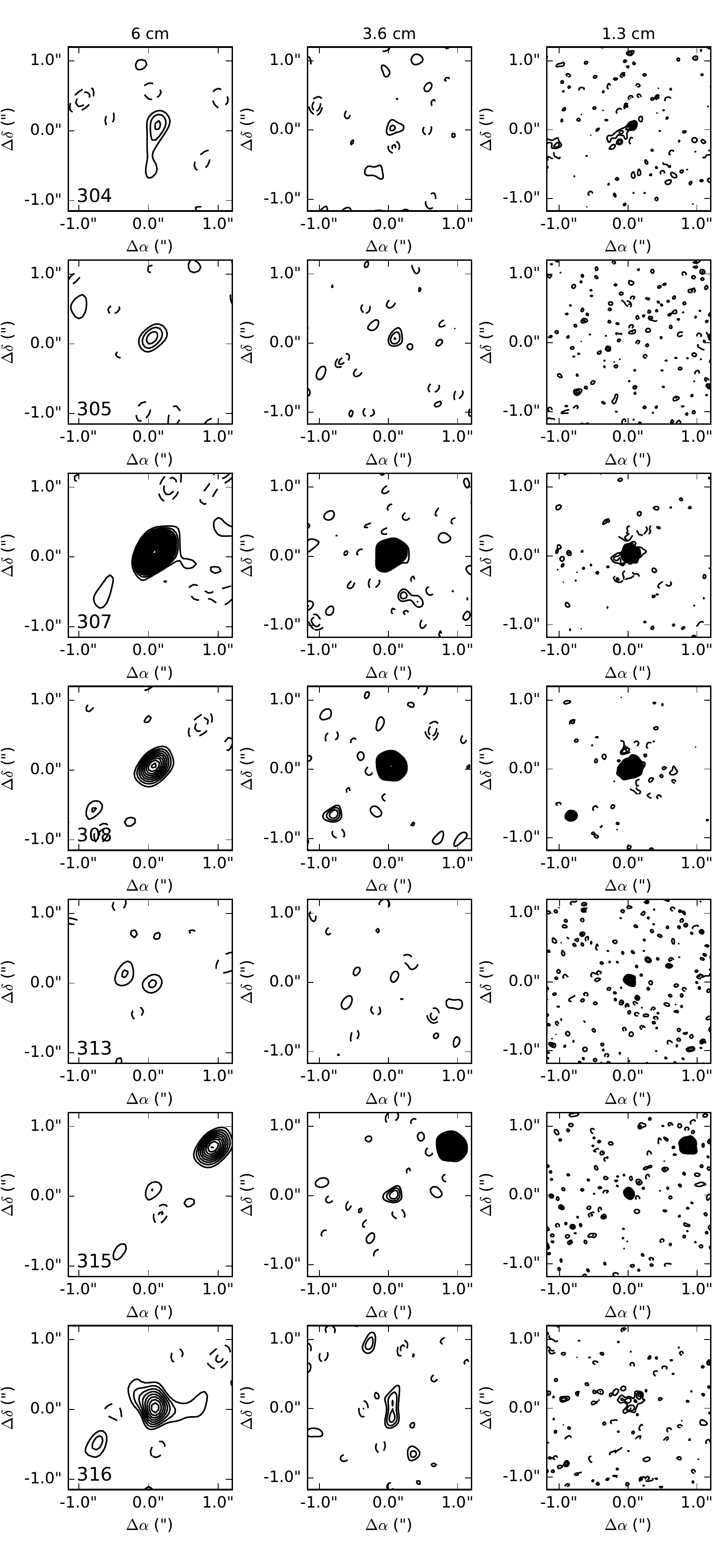}
\caption{{\it Continued}}
\end{figure*}

\begin{figure*}
\centering
\includegraphics[width=4in]{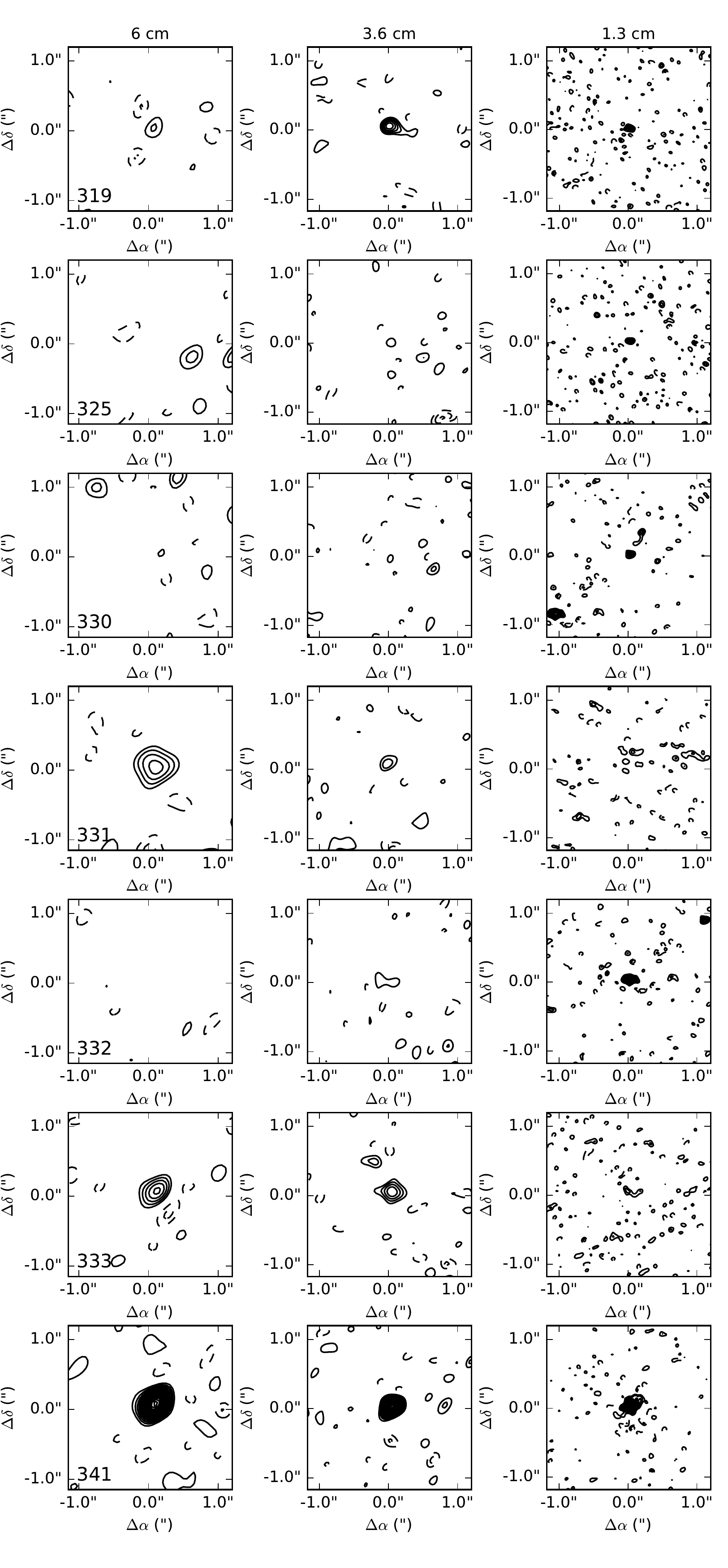}
\caption{{\it Continued}}
\end{figure*}

\begin{figure*}
\centering
\includegraphics[width=4in]{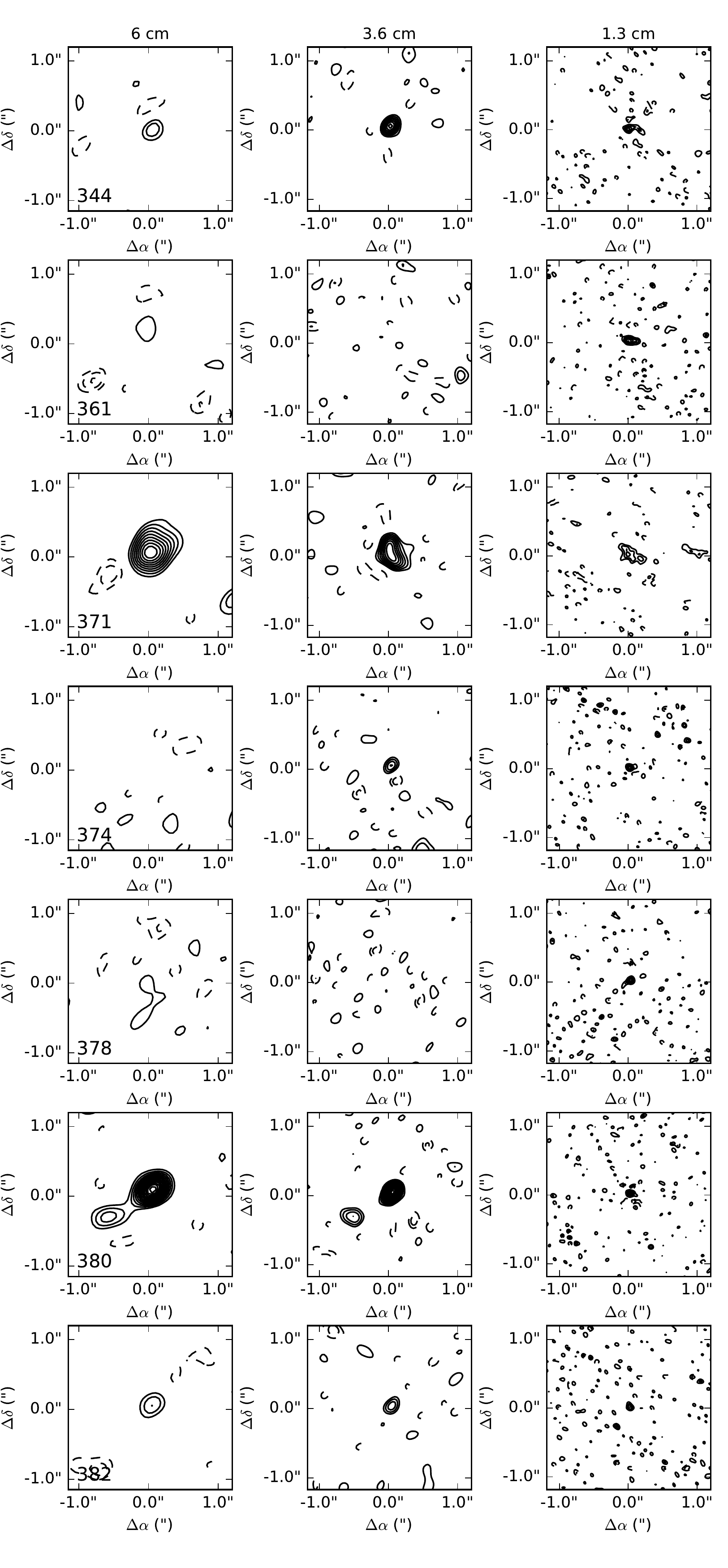}
\caption{{\it Continued}}
\end{figure*}

\begin{figure*}
\centering
\includegraphics[width=4in]{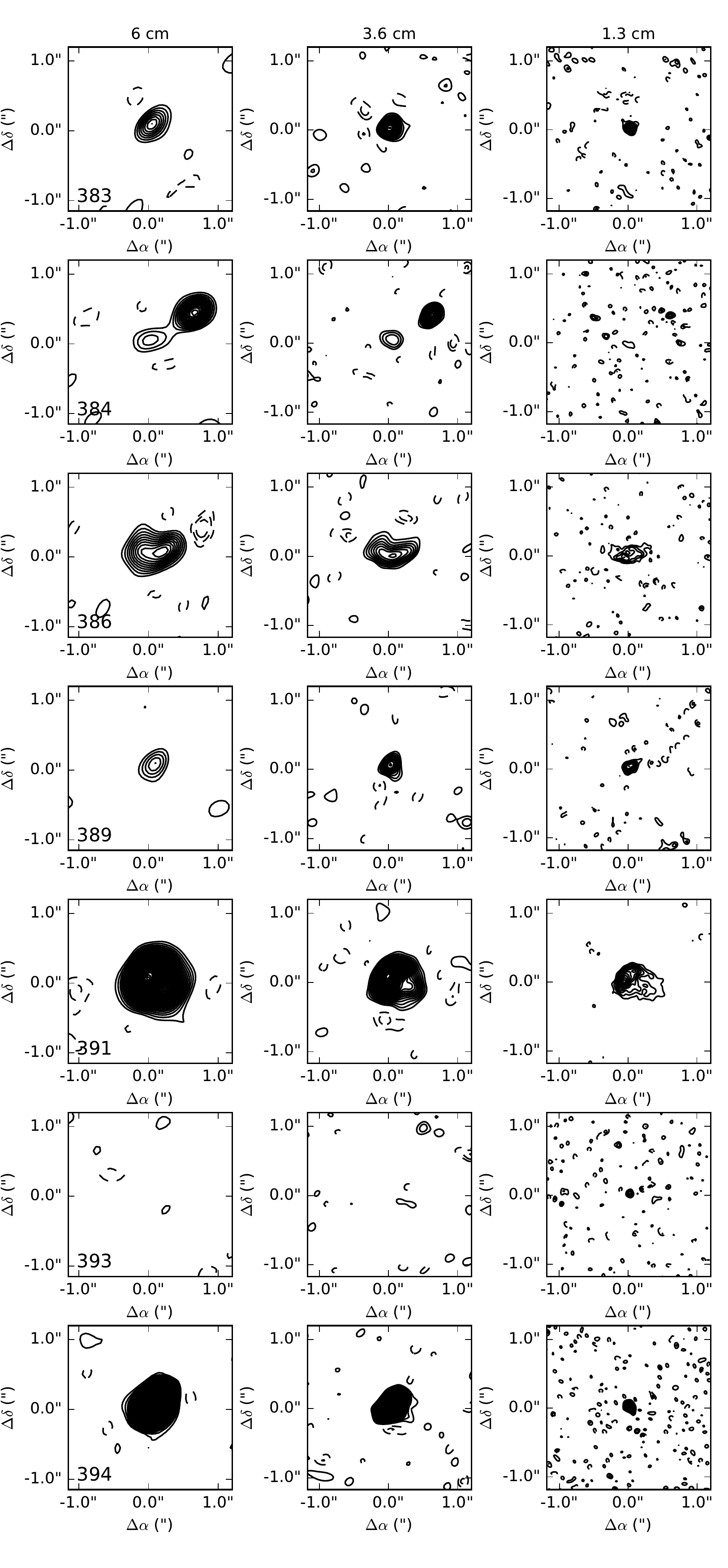}
\caption{{\it Continued}}
\end{figure*}

\begin{figure*}
\centering
\includegraphics[width=4in]{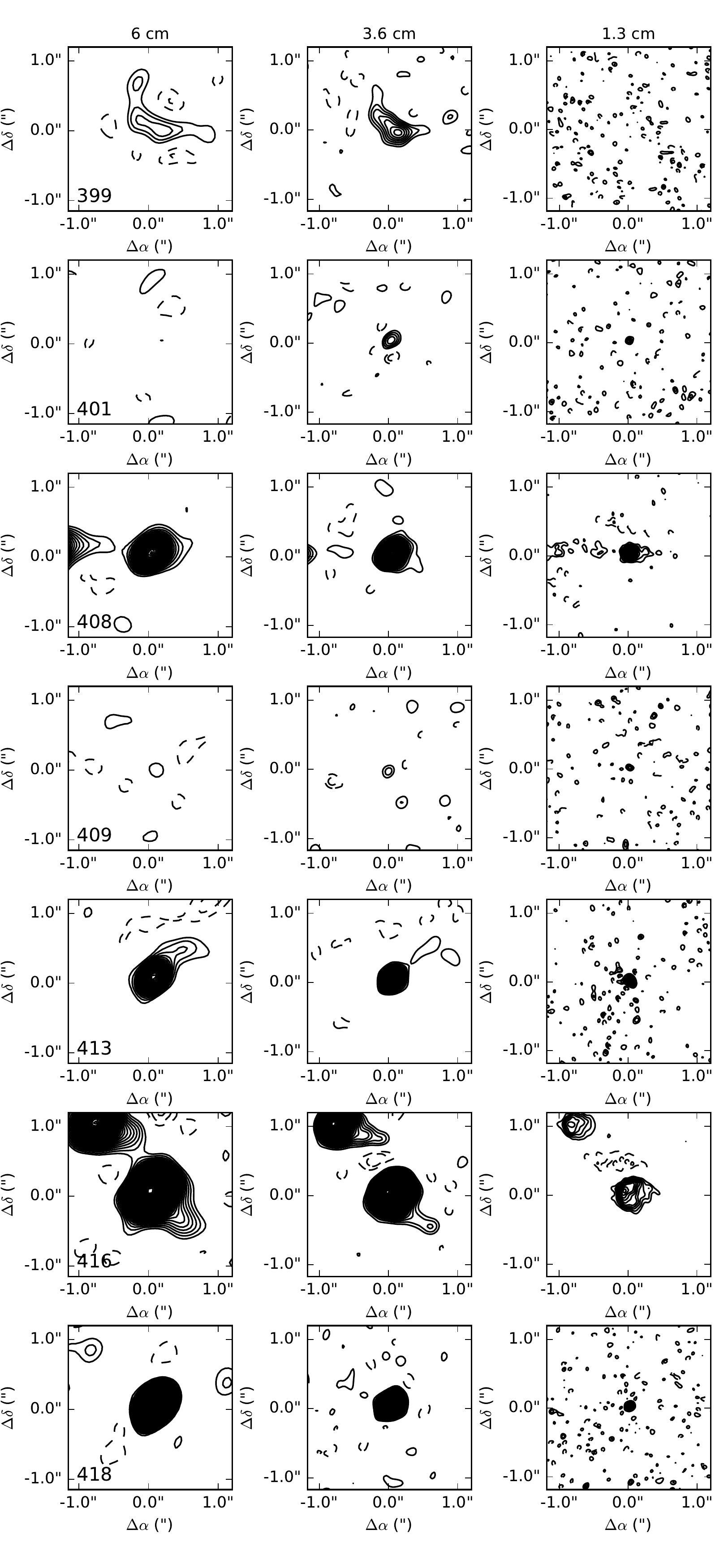}
\caption{{\it Continued}}
\end{figure*}

\begin{figure*}
\centering
\includegraphics[width=4in]{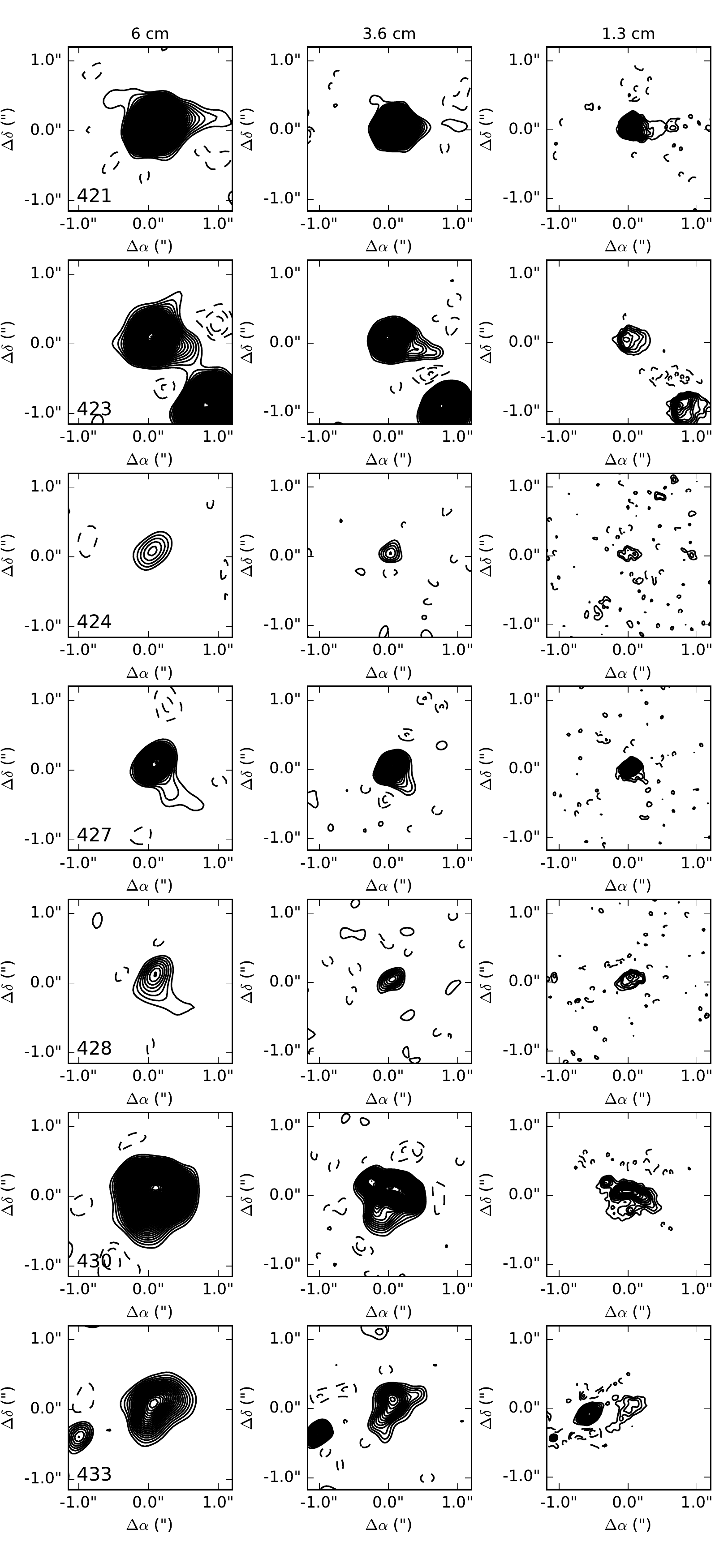}
\caption{{\it Continued}}
\end{figure*}

\begin{figure*}
\centering
\includegraphics[width=4in]{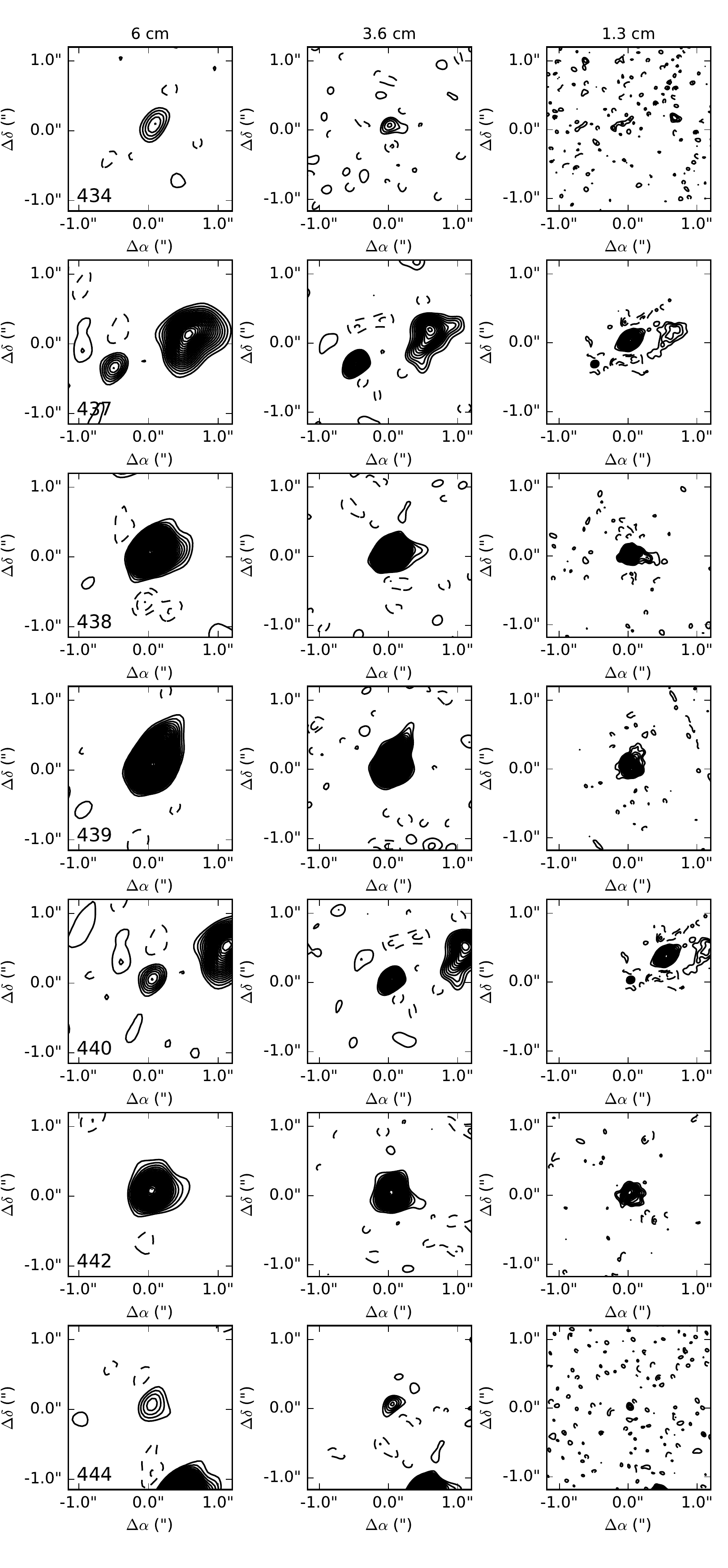}
\caption{{\it Continued}}
\end{figure*}

\begin{figure*}
\centering
\includegraphics[width=4in]{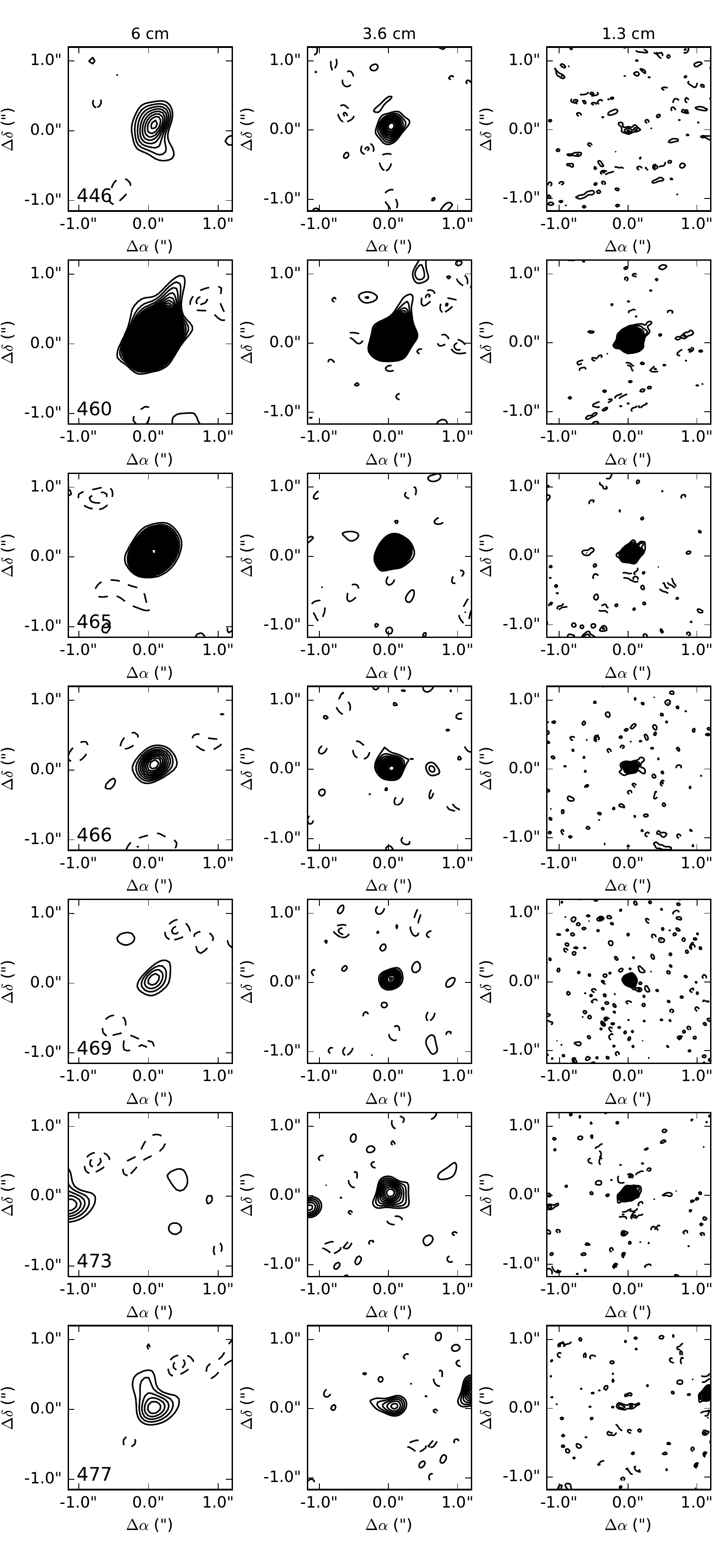}
\caption{{\it Continued}}
\end{figure*}

\begin{figure*}
\centering
\includegraphics[width=4in]{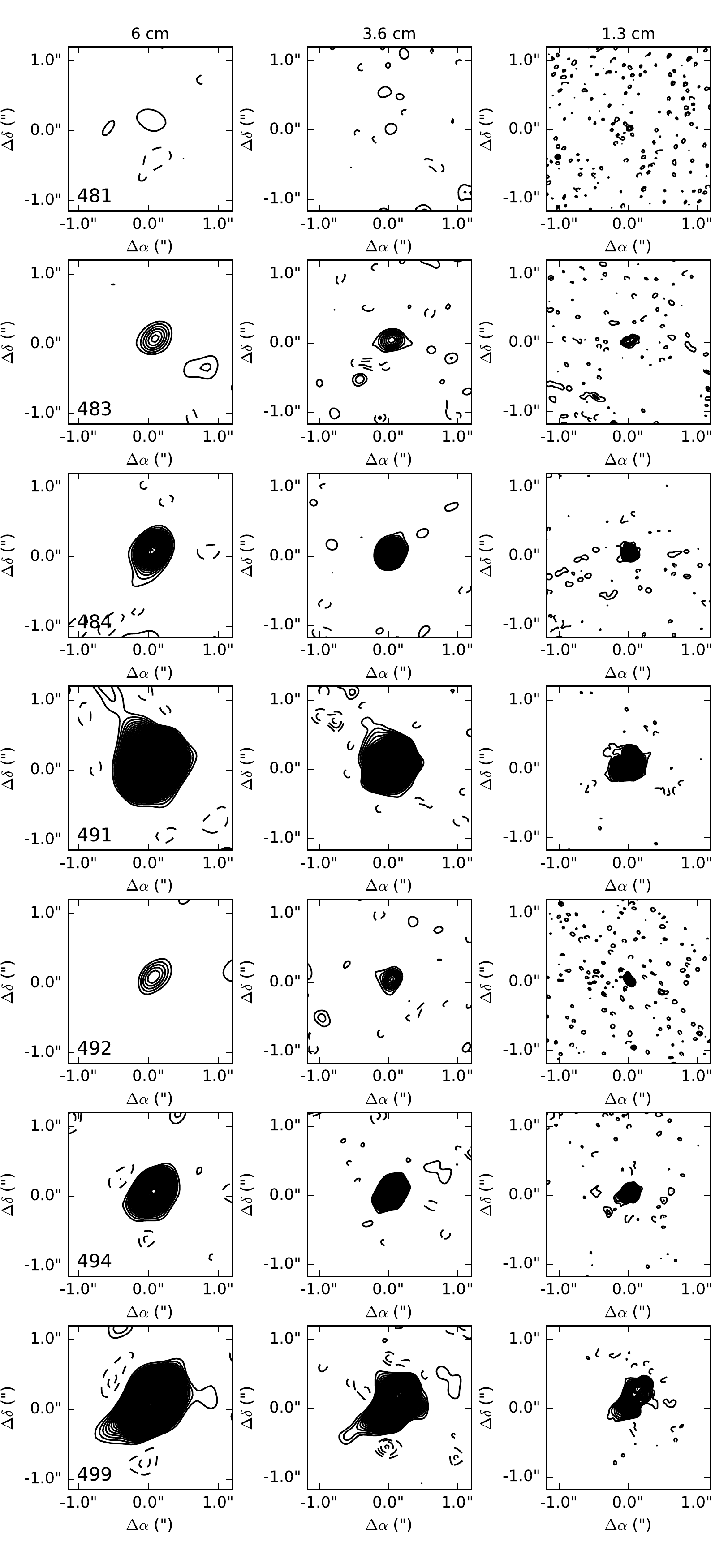}
\caption{{\it Continued}}
\end{figure*}

\begin{figure*}
\centering
\includegraphics[width=4in]{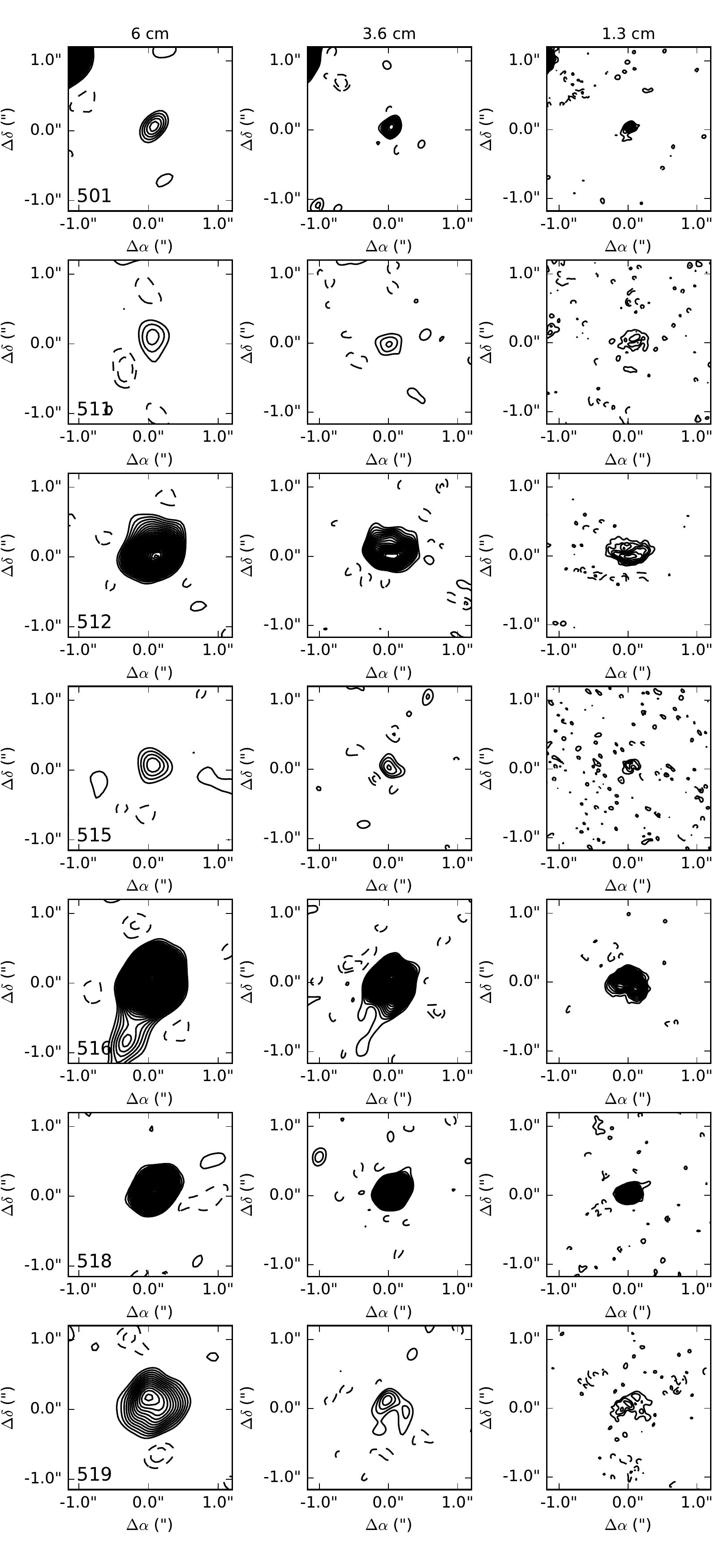}
\caption{{\it Continued}}
\end{figure*}

\begin{figure*}
\centering
\includegraphics[width=4in]{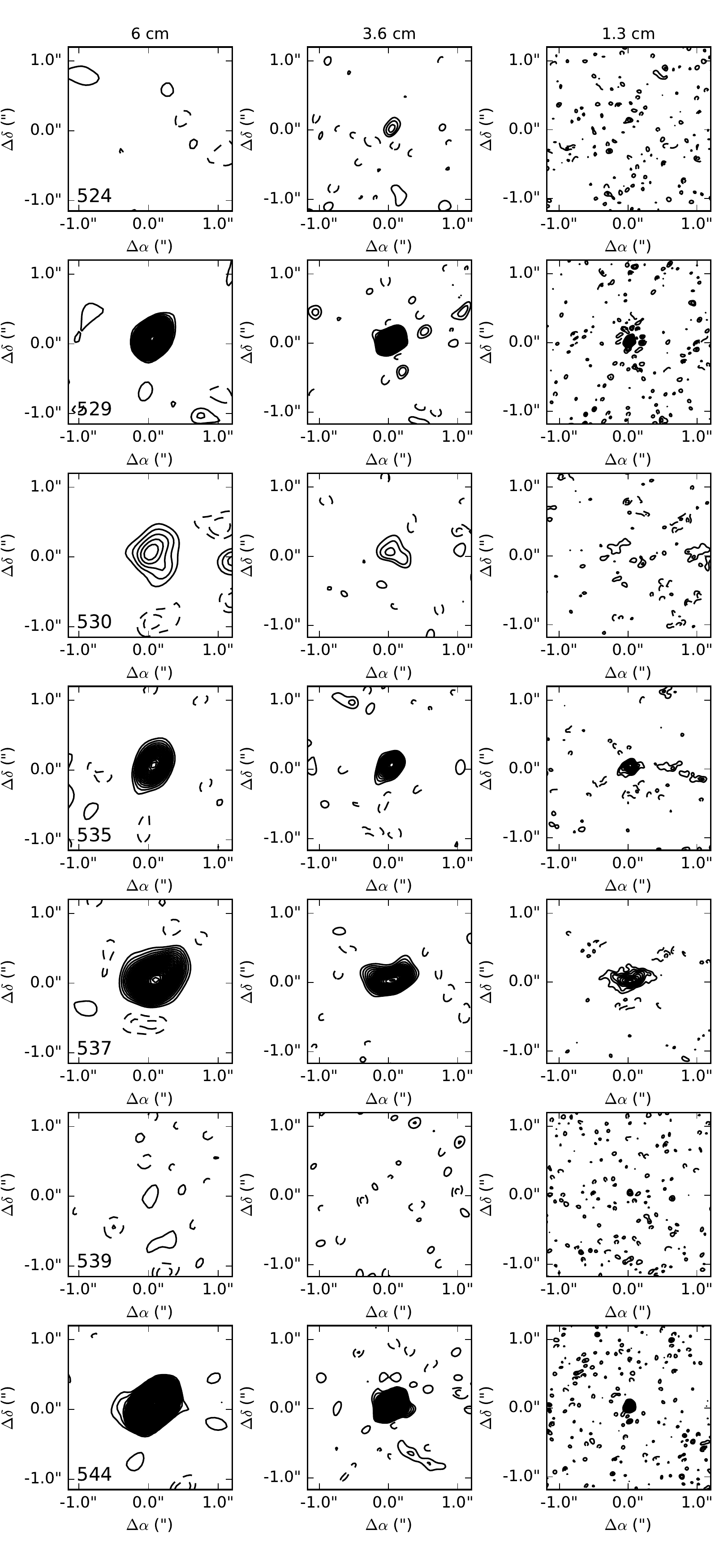}
\caption{{\it Continued}}
\end{figure*}

\begin{figure*}
\centering
\includegraphics[width=4in]{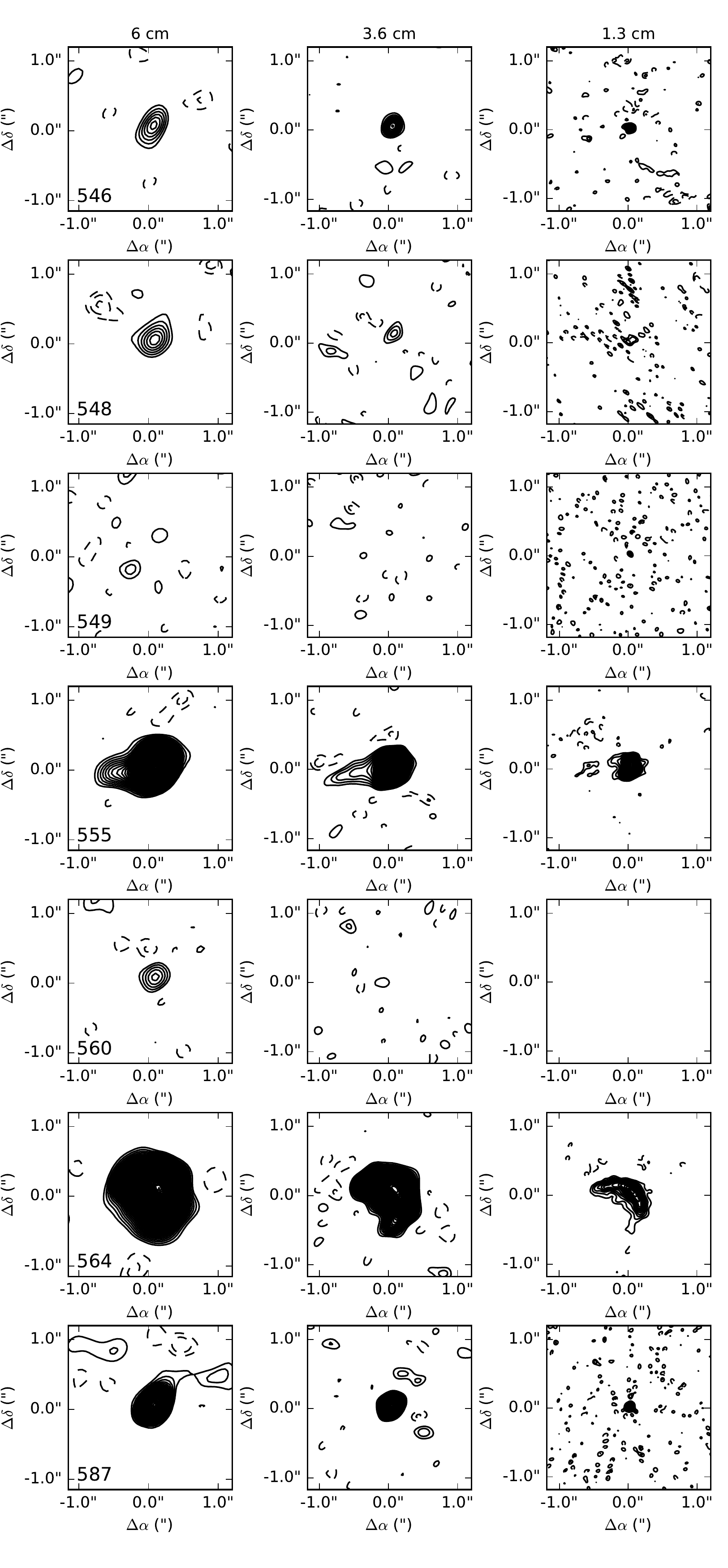}
\caption{{\it Continued}}
\end{figure*}

\begin{figure*}
\centering
\includegraphics[width=4in]{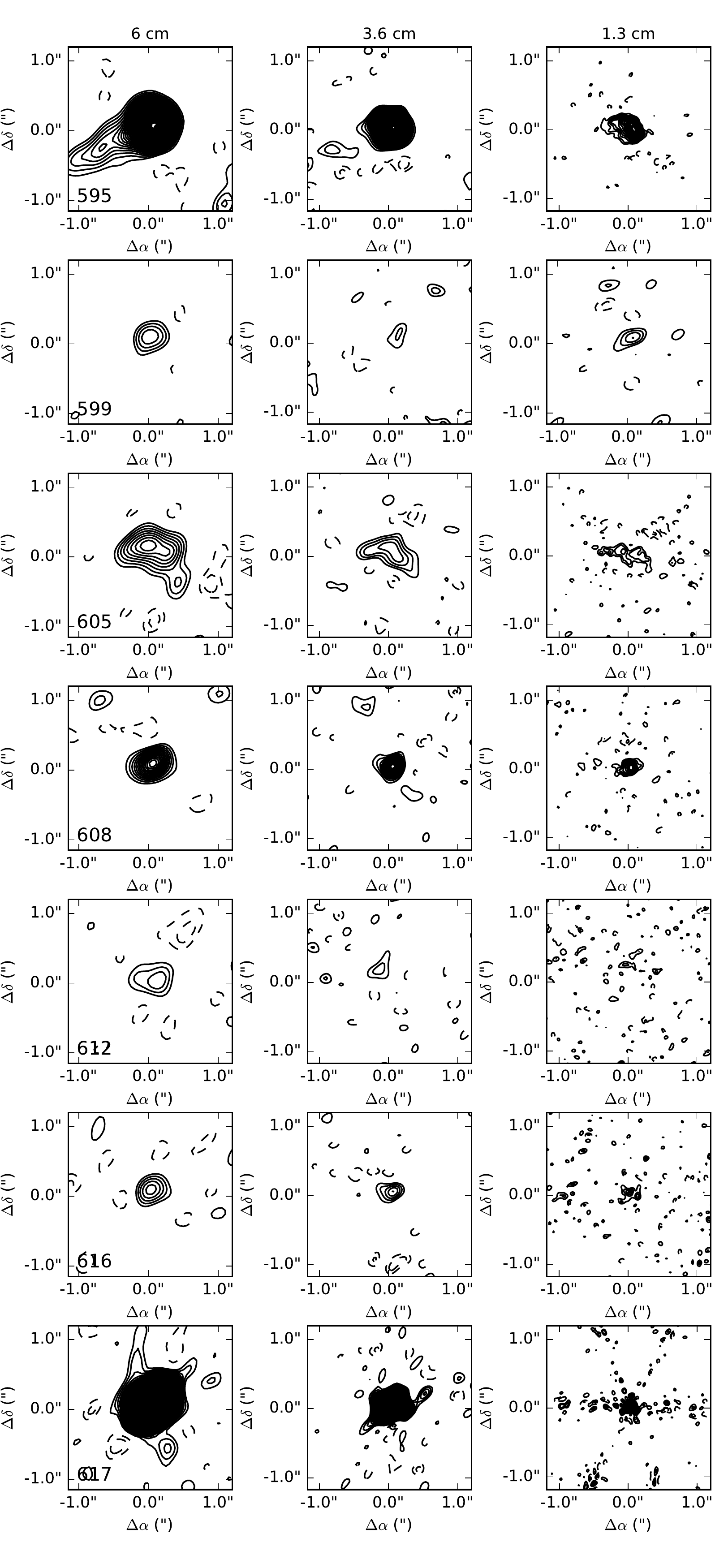}
\caption{{\it Continued}}
\end{figure*}

\begin{figure*}
\centering
\includegraphics[width=4in]{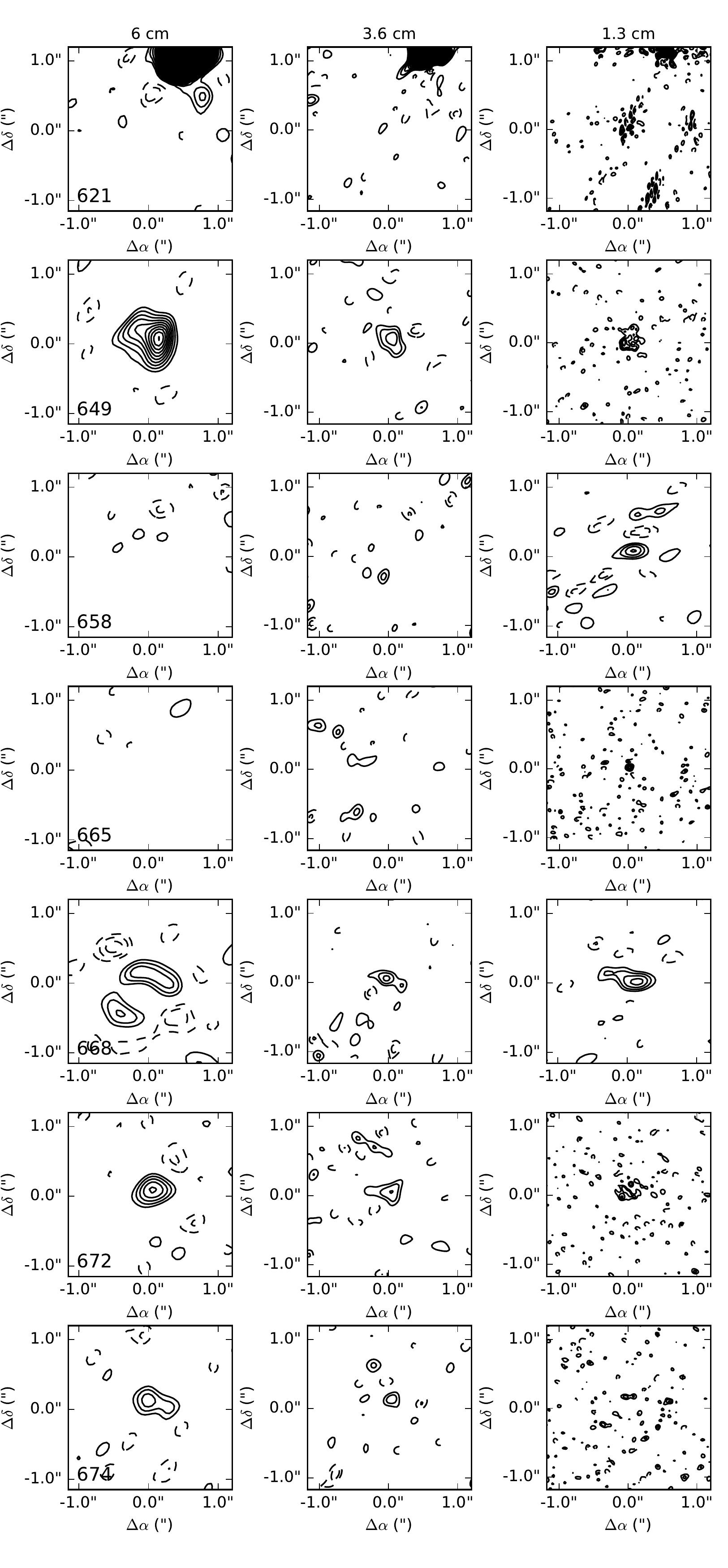}
\caption{{\it Continued}}
\end{figure*}

\begin{figure*}
\centering
\includegraphics[width=4in]{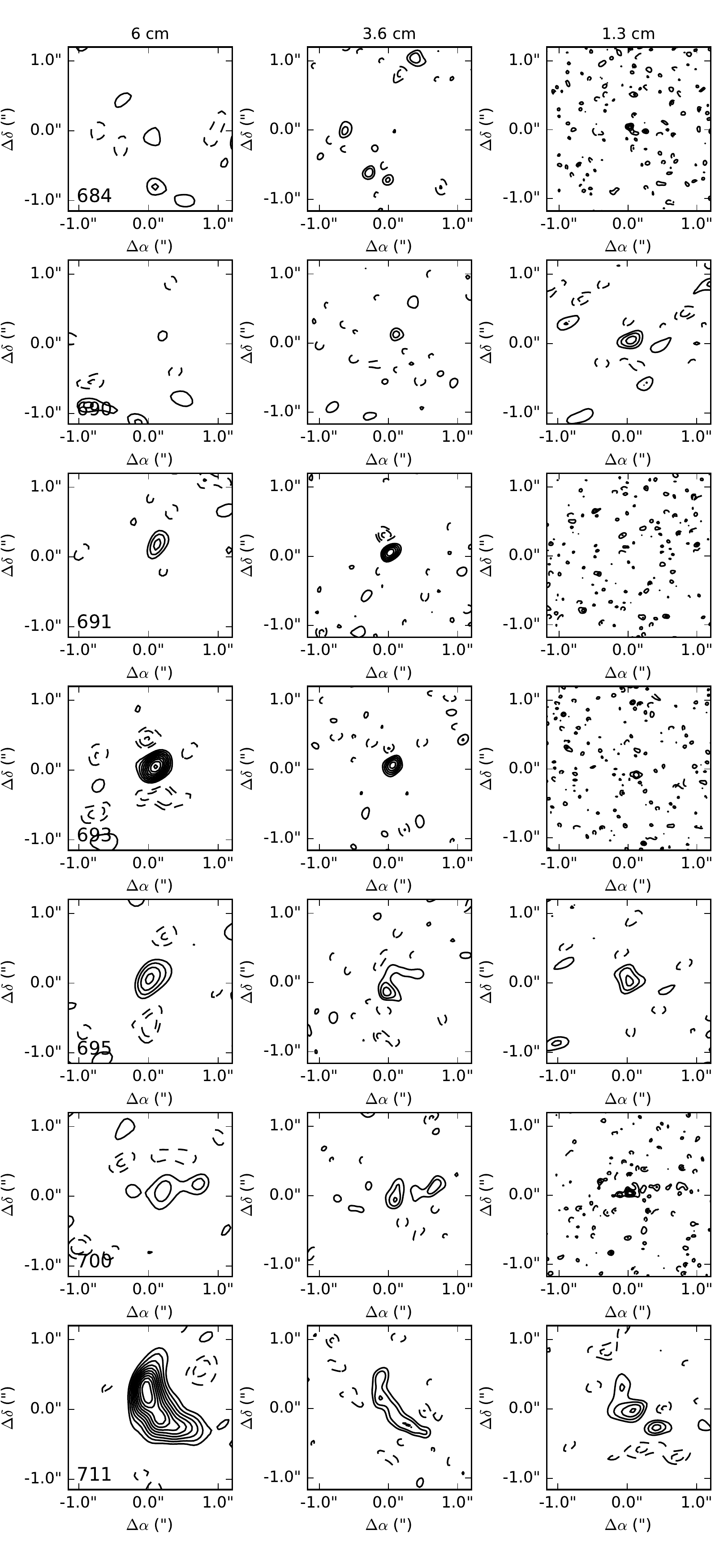}
\caption{{\it Continued}}
\end{figure*}

\begin{figure*}
\centering
\includegraphics[width=4in]{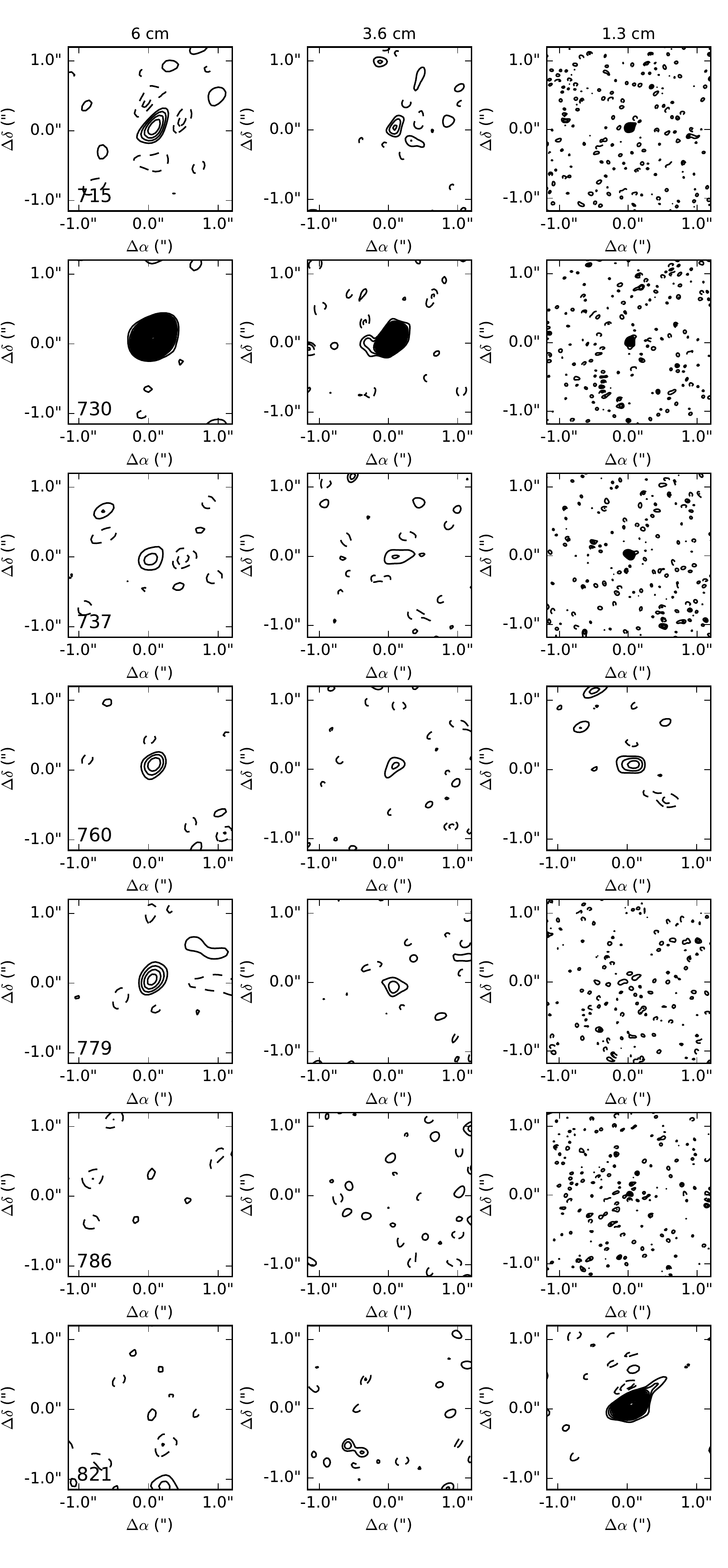}
\caption{{\it Continued}}
\end{figure*}

\begin{figure*}
\centering
\includegraphics[width=4in]{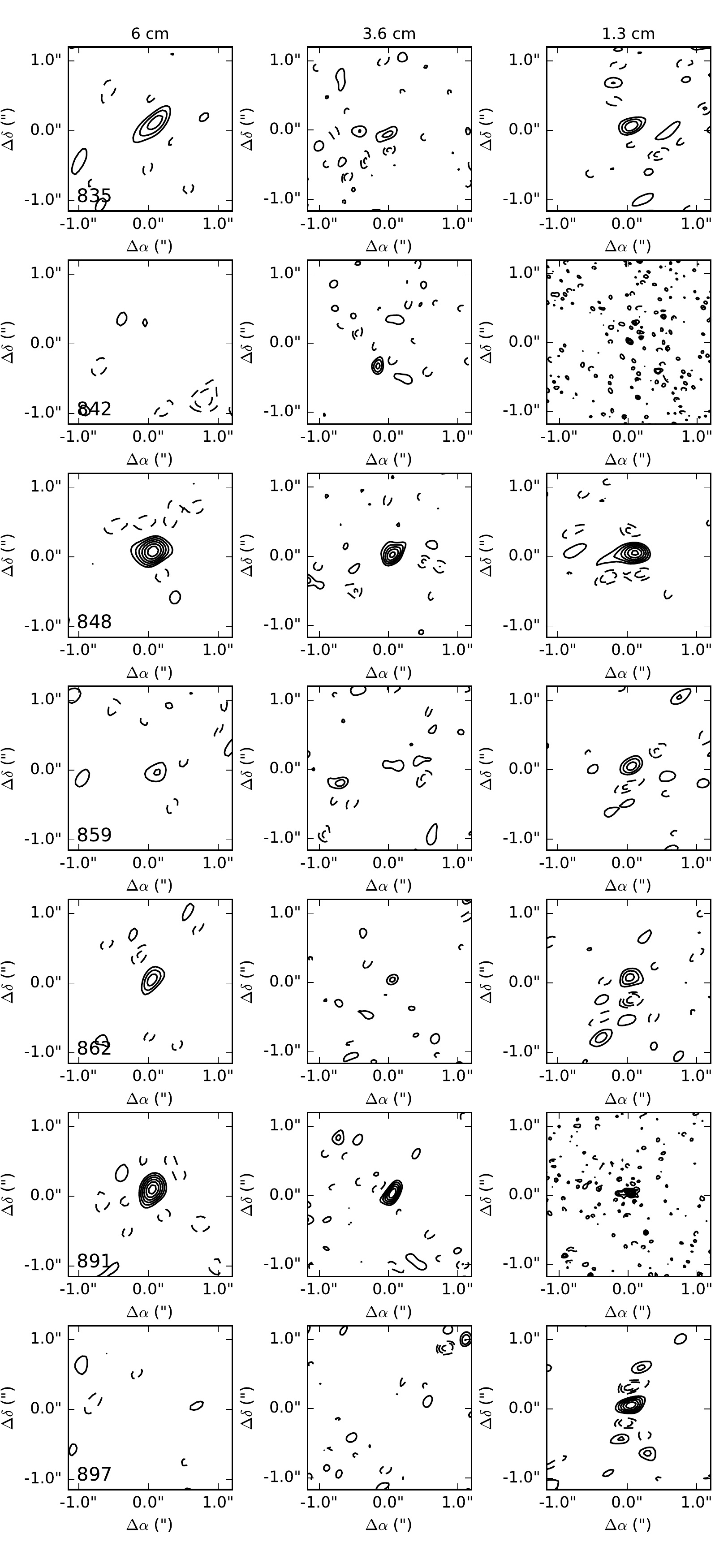}
\caption{{\it Continued}}
\end{figure*}

\begin{figure*}
\centering
\includegraphics[width=4in]{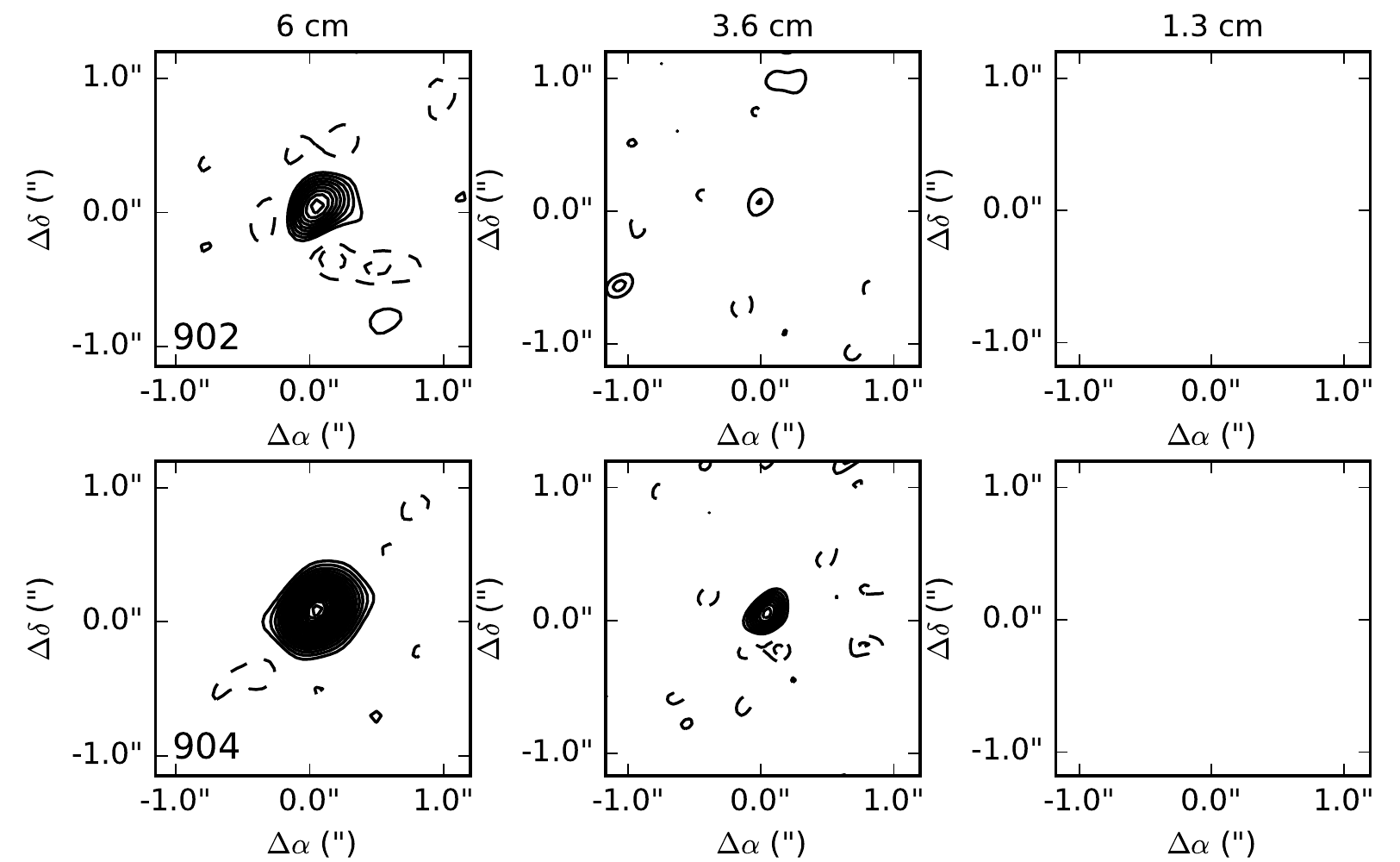}
\caption{{\it Continued}}
\end{figure*}

\begin{figure*}
\centering
\includegraphics[width=7in]{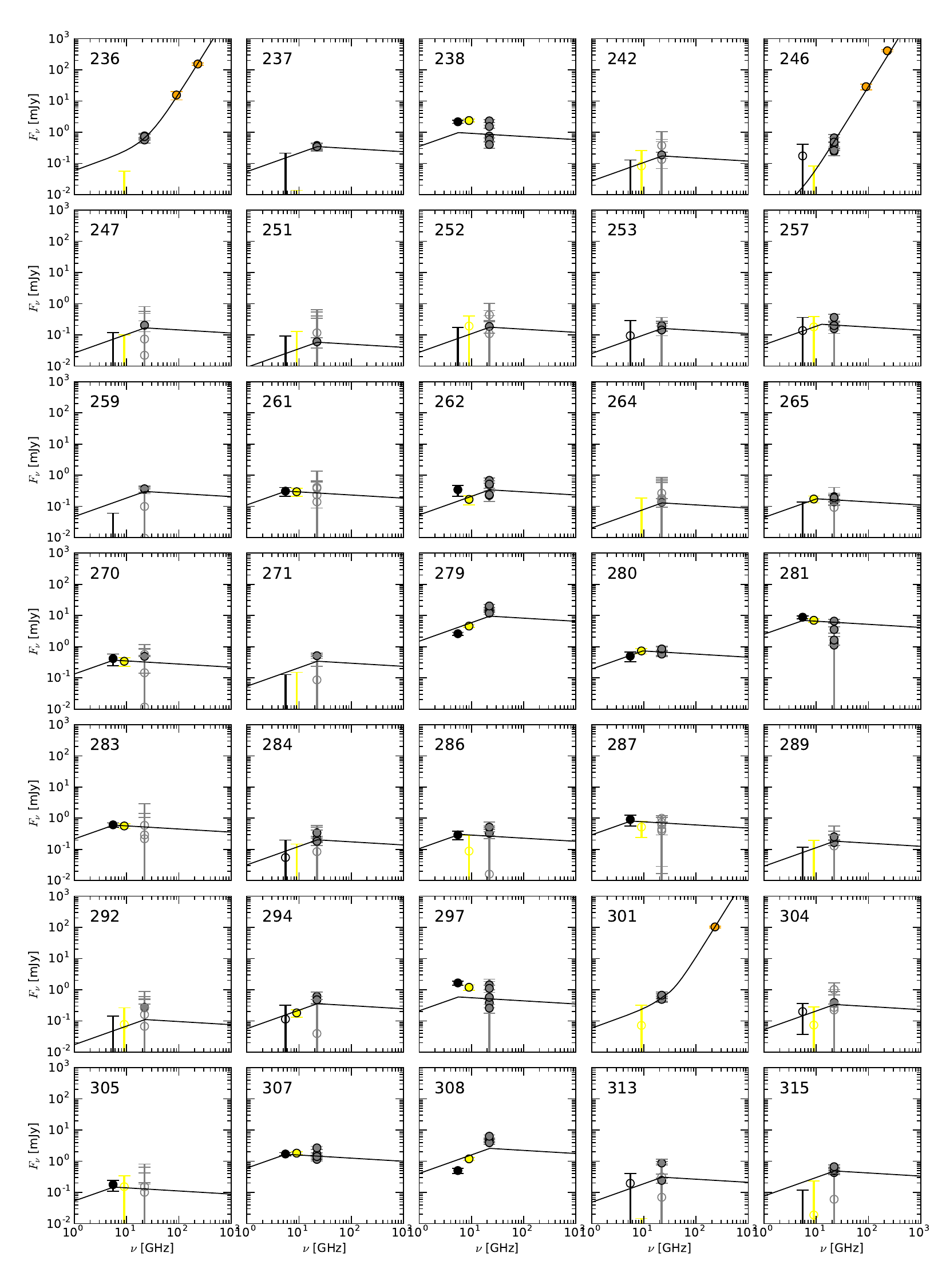}
\caption{{\it Continued}}
\end{figure*}

\begin{figure*}
\centering
\includegraphics[width=7in]{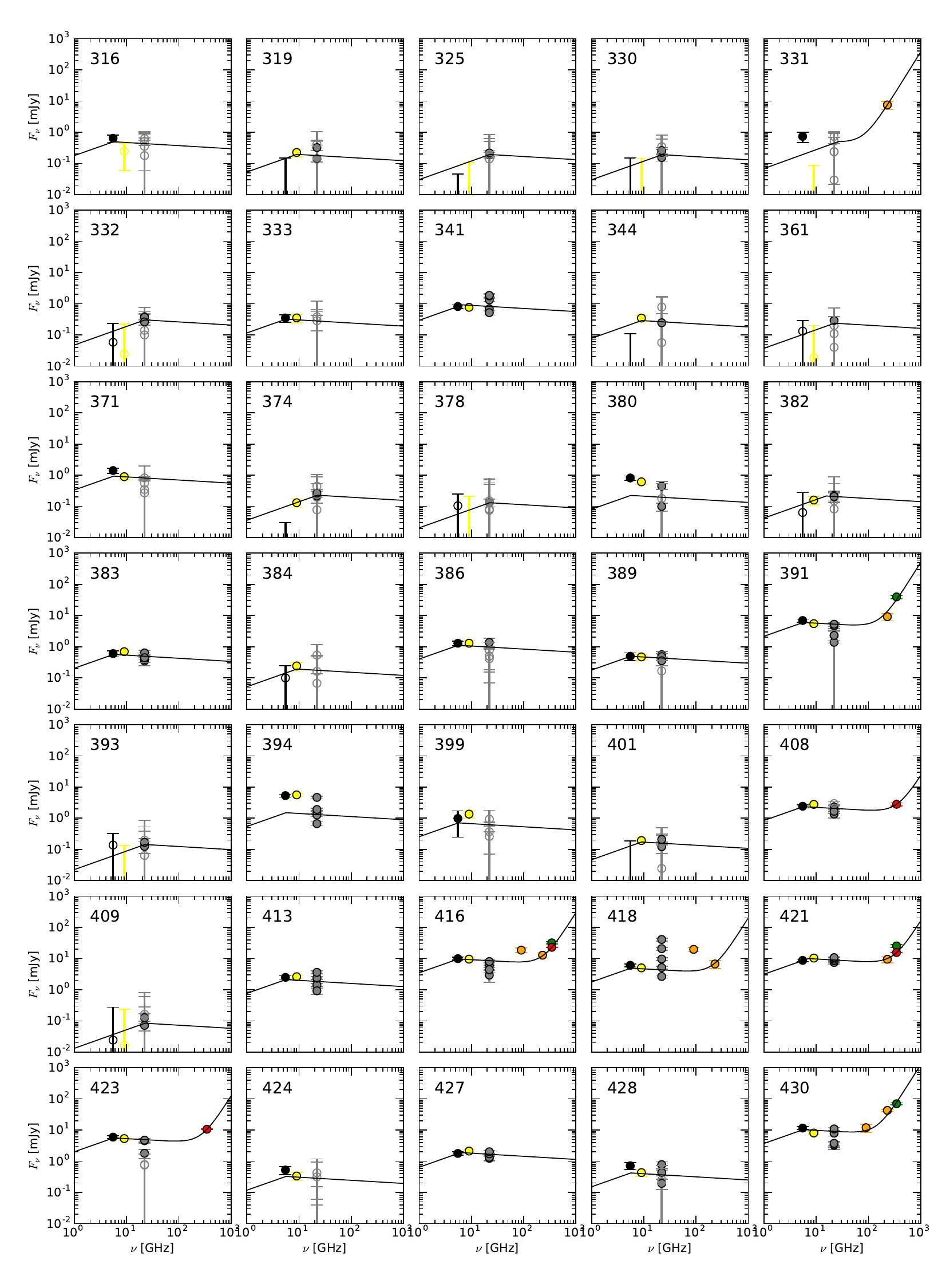}
\caption{{\it Continued}}
\end{figure*}

\begin{figure*}
\centering
\includegraphics[width=7in]{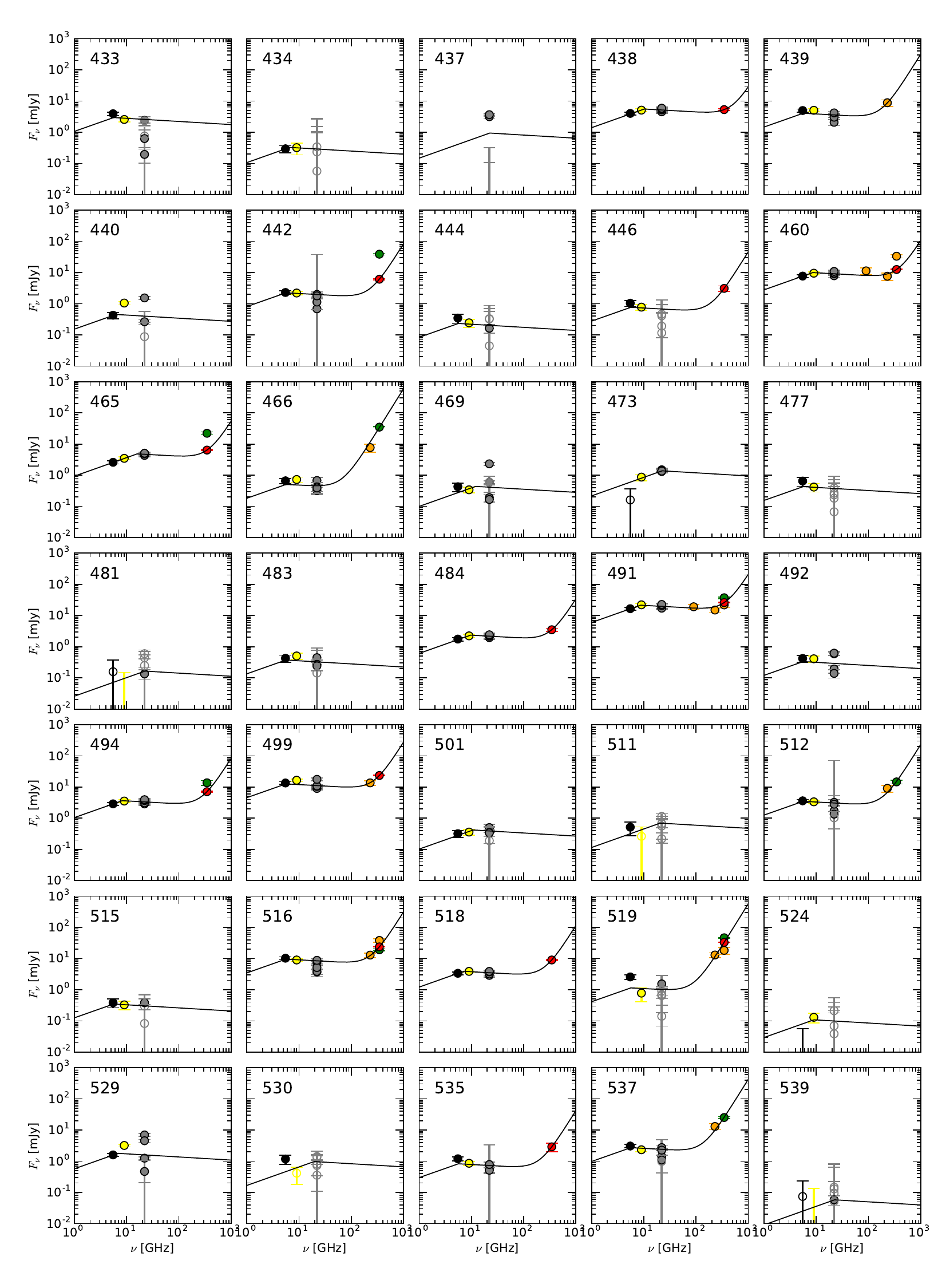}
\caption{{\it Continued}}
\end{figure*}

\begin{figure*}
\centering
\includegraphics[width=7in]{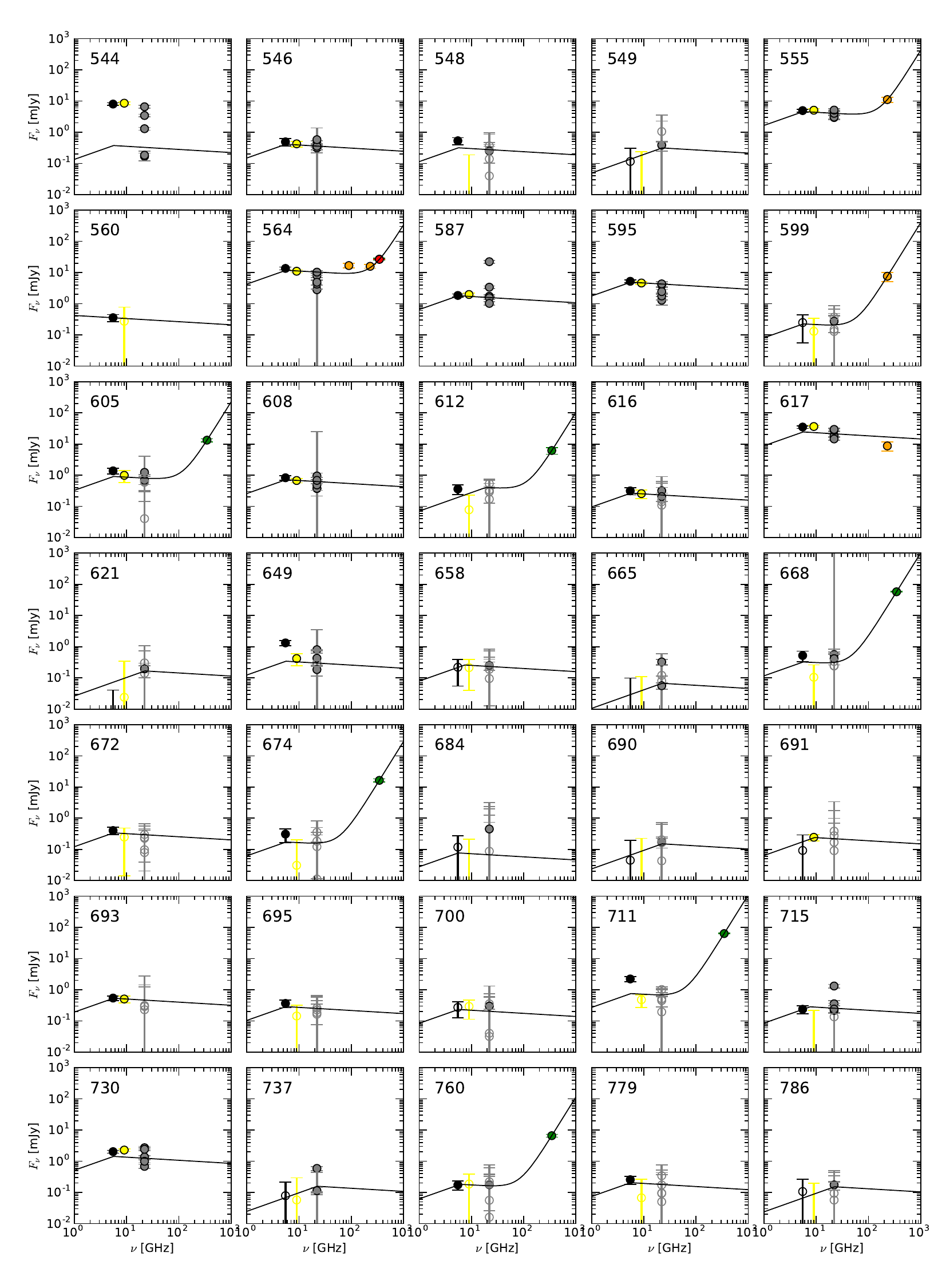}
\caption{{\it Continued}}
\end{figure*}

\begin{figure*}
\centering
\includegraphics[width=7in]{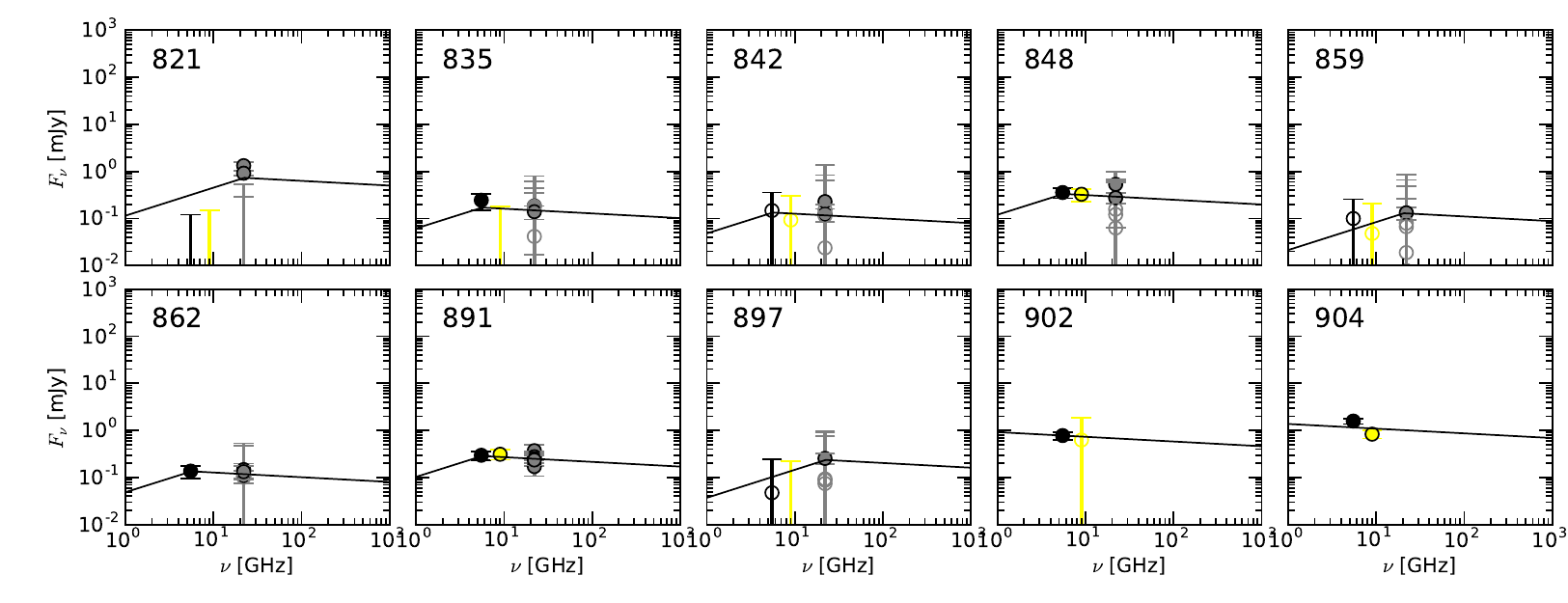}
\caption{{\it Continued}}
\end{figure*}

\end{document}